\documentclass[mnsc,nonblindrev]{informs3_hide} % current default for manuscript submission

\DoubleSpacedXI % Made default 4/4/2014 at request
\usepackage{endnotes}
\let\footnote=\endnote
\let\enotesize=\normalsize
\def\notesname{Endnotes}%
\def\makeenmark{$^{\theenmark}$}
\def\enoteformat{\rightskip0pt\leftskip0pt\parindent=1.75em
  \leavevmode\llap{\theenmark.\enskip}}
\usepackage{natbib}
 \bibpunct[, ]{(}{)}{,}{a}{}{,}%
 \def\bibfont{\small}%
\usepackage[normalem]{ulem}
\usepackage{amsmath}
\usepackage{enumitem}
\usepackage[linesnumbered,ruled,vlined]{algorithm2e}
\usepackage{algpseudocode}
\usepackage{tikz}
\usepackage{caption}
\usepackage{subcaption}
\usepackage{graphicx}
\makeatletter
\newcommand*{\rom}[1]{\expandafter\@slowromancap\romannumeral #1@}
\makeatother

\TheoremsNumberedThrough     % Preferred (Theorem 1, Lemma 1, Theorem 2)
\ECRepeatTheorems

\EquationsNumberedThrough    % Default: (1), (2), ...
\newcommand{\RR}{\mathbb{R}}
\newcommand{\ZZ}{\mathbb{Z}}
\newcommand{\EE}{\mathbb{E}}
\newcommand{\PP}{\mathbb{P}}
\newcommand{\Gap}{\textbf{Gap}}
\newcommand{\ARO}{\textup{ARO}}
\newcommand{\LP}{\textup{LP}}
\allowdisplaybreaks % allow page breaks, if needed

\begin{document}
%%%%%%%%%%%%%%%%

%\setcounter{page}{1}
%\DOI{10.1287/xxxx.0000.0000}%

\RUNAUTHOR{Chen et al.}

\RUNTITLE{Assortment Optimization Under MVMNL}

% \TITLE{Bundling Information Goods of Decreasing Value}
\TITLE{Assortment Optimization Under the Multivariate MNL Model\footnote{Authors are listed in alphabetical order.}}%

\ARTICLEAUTHORS{%
\AUTHOR{Xin Chen}
\AFF{H. Milton Stewart School of Industrial and Systems Engineering, Georgia Institute of Technology, Atlanta, Georgia, \EMAIL{xin.chen@isye.gatech.edu}}
\AUTHOR{Jiachun Li}
\AFF{Tsinghua University, Beijing, China,  \EMAIL{lijc19@mails.tsinghua.edu.cn}}
\AUTHOR{Menglong Li}
\AFF{Department of Management Sciences, City University of Hong Kong, Hong Kong,  \EMAIL{mengloli@cityu.edu.hk}}
\AUTHOR{Tiancheng Zhao\footnote{Corresponding author}}
\AFF{Gies College of Business, University of Illinois at Urbana-Champaign, Urbana, Illinois,  \EMAIL{tz14@illinois.edu}}
\AUTHOR{Yuan Zhou}
\AFF{Yau Mathematical Sciences Center, Tsinghua University, Beijing, China,  \EMAIL{yuan-zhou@tsinghua.edu.cn}}
} % end of the block

\ABSTRACT{%
We study an assortment optimization problem under a multi-purchase choice model in which customers choose a bundle of up to one product from each of two product categories. Different bundles have different utilities and the bundle price is the summation of the prices of products in it. For the uncapacitated setting where any set of products can be offered, we prove that this problem is strongly NP-hard. We show that an adjusted-revenue-ordered assortment provides a $\frac{1}{2}$-approximation. Furthermore, we develop an approximation framework based on a linear programming relaxation of the problem and obtain a 0.74-approximation algorithm. This approximation ratio almost matches the integrality gap of the linear program, which is proven to be at most 0.75. For the capacitated setting, we prove that there does not exist a constant-factor approximation algorithm assuming the Exponential Time Hypothesis. The same hardness result holds for settings with general bundle prices or more than two categories. Finally, we conduct numerical experiments on randomly generated problem instances. The average approximation ratios of our algorithms are over 99\%.
}%


%\HISTORY{This paper was
%first submitted on April 12, 1922 and has been with the authors for
%83 years for 65 revisions.}

\maketitle
%%%%%%%%%%%%%%%%%%%%%%%%%%%%%%%%%%%%%%%%%%%%%%%%%%%%%%%%%%%%%%%%%%%%%%

% Samples of sectioning (and labeling) in OPRE
% NOTE: (1) \section and \subsection do NOT end with a period
%       (2) \subsubsection and lower need end punctuation
%       (3) capitalization is as shown (title style).
%
%\section{Introduction.}\label{intro} %%1.
%\subsection{Duality and the Classical EOQ Problem.}\label{class-EOQ} %% 1.1.
%\subsection{Outline.}\label{outline1} %% 1.2.
%\subsubsection{Cyclic Schedules for the General Deterministic SMDP.}
%  \label{cyclic-schedules} %% 1.2.1
%\section{Problem Description.}\label{problemdescription} %% 2.

% Text of your paper here

\section{Introduction}\label{sec:intro}

Assortment optimization, which aims to maximize revenue or profit by selecting a combination of products from a product universe, is of fundamental importance in revenue management both theoretically and practically.  Traditional assortment optimization problems typically assume that customers can only purchase one item at a time. In other words, the model leads to \textit{single-choice} outcome. Although this assumption is valid in some scenarios, there are  many scenarios in which the customer may want to purchase multiple items as a bundle, i.e., a \textit{multi-choice} outcome. For example, when purchasing different pasta shapes and pasta sauce types, the customers may consider that different pasta shapes are suitable for different pasta sauces (e.g., angel hair pasta is best for tomato-based sauces). Thus, there exists complementarity between different product pairs from these two categories in addition to substitutability between products of the same category. In addition, the customers can choose to purchase a bundle of pasta and sauce or just one of them. Another example is when customers purchase a bundle of PC components, including CPU, memory, motherboards, etc. When making the purchase decision, customers may need to consider the compatibility between these parts. If the two parts do not work together, customers will not likely choose that bundle. In addition, customers may want to choose the parts of a similar level of performance such that no lower-end component limits the overall performance of the PC system.

The above two examples reveal two features of customers' purchasing behaviors in the presence of multiple categories. One is that customers often consider the assortment of one category when making decisions for another category. The other is that complicated interactions exist between products in a bundle that determines the utility of the bundle to customers. The interaction might include complementarity between products and compatibility between products. To account for these features, some multi-choice models are proposed in marketing and economics literature. One of the most popular models in this stream of literature is the Multivariate MNL (MVMNL) model, where ``Multivariate'' stands for multi-choice, in contrast with the classical single-choice (univariate) MNL model. In the MVMNL model, customers are presented with multiple product categories and can choose at most one product from each category, to form a bundle. Each product bundle is assigned a utility parameter, which may differ from the sum of the utilities of the individual products inside the bundle. The customer chooses the utility-maximizing product bundles in a way similar to choosing the utility-maximizing product in the univariate MNL model. We redirect the readers to Section 2 for a more detailed review of the literature on multi-choice models.

Although the \textit{multi-choice} models have received much attention in the marketing and economics literature, the effort has been focused on modeling the utility of bundles. Little has been done on developing a fundamental understanding of the resulting assortment optimization problems. We review few attempts on assortment optimization under multi-choice models in Section \ref{sec:LR}. This paper is one of the first attempts to fill this gap. Specifically, in this paper, we focus on a MVMNL model with two product categories in which the customers are allowed to purchase a bundle of at most one product from each category. 
The price of a bundle is the summation of the prices of products in it, and the utility of a bundle can have arbitrary values. We refer to the two-category MVMNL model with these assumptions as the \textit{base model}. 
We provide a strong NP-hardness result for the unconstrained assortment optimization problem and a simple 0.5-approximation algorithm by considering adjusted-revenue-ordered assortments (an analogue of revenue-ordered assortments). Moreover, we propose an approximation framework based on an LP relaxation of the problem and improve the approximation ratio to 0.74. We show that this approximation ratio is within 0.01 of the integrality gap of the LP relaxation. Extensive numerical experiments are conducted and demonstrate that the approximation ratios of our algorithms are much better than the theoretical approximation guarantees. We also prove that there does not exist a constant-factor approximation algorithm for various extensions including the capacitated problem, the two-category MVMNL model with more general bundle price structures, and MVMNL models with more than two categories.  
% We show that the assortment optimization problem is APX-hard, which rules out a PTAS for this problem. Motivated by the hardness result, we develop polynomial-time approximation algorithms based on the LP relaxation of the problem that achieves a constant approximation guarantee. We also show that extending this problem into some more general settings will result in an assortment optimization problem that is hard to approximate within any constant ratio.
%We note that considering arbitrary product interactions results in a model that is considerably harder than that studied in the literature. For example, \cite{lyu2021assortment} develop FPTAS for assortment optimization under the multi-category multivariate MNL model for group-based product interactions, while our problem is APX-hard even for two categories, for which no PTAS exists.  (follow the suggested flow?)

\subsection{Technical challenges and contributions}
We summarize our major results below. %{\color{red}may need some edits and enrichment}

\textbf{Strong NP-hardness result for the base model.} 
% We study the assortment optimization problem under the two-category MVMNL model. In our model, the utility of product bundles can be determined by arbitrary interaction between products in the bundle, and the price of bundles is the summation of individual prices. As we will show in the literature review in Section \ref{sec:LR}, there can be multiple types of interactions, which does not lead to an easy formula to compute bundles' utility. 
We show that even the unconstrained assortment optimization problem under the base model is strongly NP-hard, which rules out the existence of FPTAS for the problem. This is proved by a reduction from a strongly NP-hard problem, \textsc{Max-DiCut} on a directed acyclic graph. 
% We note that APX-hardness results are rarely seen in the OR/OM literature. 
This result implies that our problem is harder to approximate than some closely related problems in the literature, e.g., the assortment optimization problems studied in \cite{lyu2021assortment} for an MVMNL model with special bundle utilities that admits FPTAS. 
%and  \cite{ghuge2022constrained} for the paired combinatorial model  both admit PTAS, i.e., it can be approximated with any ratio close to one.

\textbf{0.5-approximation of adjusted-revenue-ordered assortments.} We propose an analogue of revenue-ordered assortments, termed \textit{adjusted-revenue-ordered} assortment, and show that the one with the highest revenue is a 0.5-approximation. The \textit{adjusted-revenue-ordered} assortments are constructed by fixing a revenue-ordered assortment of one category and selecting a level set of the products from the other category based on the order of certain weighted revenue. 
% However, as shown both by the theoretical approximation guarantee and the numerical experiments, the performance of this simple algorithm leaves a lot of room for improvement.
However, the worst-case approximation ratio of \textit{adjusted-revenue-ordered} assortments is 0.5.

\textbf{Partition-and-optimize approximation framework.} To better approximate the assortment optimization problem under the base model, we propose an approximation framework based on an LP relaxation of the problem, which we termed \textit{partition-and-optimize} approximation framework. The LP relaxation is obtained by the McCormick inequalities and its optimal solution is shown to be $\{0,\frac{1}{2},1\}$-valued (after scaling). 
The framework has three steps after solving the LP relaxation and getting an non-integral optimal solution. First, we \textit{partition} products with half-integral solutions in both categories into $K$ level sets, defined by some cutoff points, and construct $K$ different integral solutions using combinations of different level sets for the two categories.
Second, we use a quadratic program to compute the approximation ratio of the $K$ integral solutions given the cutoff points of the level sets.  
Third, the cutoff points of the level sets are \textit{optimized} to get the highest approximation ratio.  

We prove that the approximation framework with $K=6$ has an approximation guarantee of at least 0.74. This approximation ratio almost closes the integrality gap of the LP relaxation, which is shown to be at most 0.75. 
% To the best of our knowledge, this result is the first constant-factor approximation to the MVMNL model with arbitrary bundle utilities (or product interactions in a bundle). 
Moreover, our numerical experiments demonstrate that the actual approximation ratio of our algorithm is much better than the theoretical guarantee.

\textbf{$(\text{Integrality gap}-\varepsilon)$-approximation scheme.} 
We propose an approximation scheme for the base model that outputs an assortment with an approximation factor within $\varepsilon$ of the integrality gap of the LP relaxation for any positive $\varepsilon$ but with a time complexity depending exponentially on $\varepsilon$. Similar to the partition-and-optimize approximation framework, we first partition the products in each category into $K=1/\varepsilon$ blocks, but with a set of thresholds equally spread between $0$ and $1$. We then find the best assortment among all possible combinations of the blocks. We show that this assortment achieves an approximation ratio no less than the integrality gap minus $\varepsilon$. Moreover, this algorithm can be easily extended to other interesting cases of bundle price structures. For example, the price of the bundle can be certain functions of the summation of the prices of products in it. This allows us to model certain discount schemes for bundles, e.g., when the bundle price is the sum of the individual prices subtracted by a fixed discount or multiplied by a fixed discount factor.

% We demonstrate how to extend the \textit{partition-and-optimize} framework to achieve an approximation ratio arbitrarily close to the integrality gap. Similar to the approximation framework, we first partition the products in each category into $K=1/\varepsilon$ level sets, with a set of thresholds equally spread among $0$ and $1$. Since we do not optimize the set of thresholds, we find the best assortment among all possible combinations of the level sets instead of considering just $K$ combinations of different level sets. We show that this approximation algorithm can achieve an approximation ratio arbitrarily close to the integrality gap as $\varepsilon$ goes to $0$.

\textbf{Extensions to more general settings.} In addition to the base model, we investigate the assortment optimization problem under more general settings including the base model with cardinality constraints, the two-category MVMNL model with arbitrary bundle price structures, and the three-category MVMNL model. We prove that there does not exist a constant-factor approximation for these extensions assuming the widely-used \textit{Exponential Time Hypothesis (ETH)} in theoretical computer science community. These results further illustrate the hardness of our problem under the base model in view of the existence of constant-factor approximation for extensions of some closely related problems in the literature (e.g., the constant-factor approximation for the cardinality-constrained assortment optimization problems under the PCL model studied in \cite{ghuge2022constrained}).

\subsection{Notations and Terminologies} 
We list the notations and terminologies used throughout this paper. The real space and integer space are denoted by $\RR$ and $\ZZ$, respectively. Define $\RR_+=\{x\in \RR\mid x\geq 0\}$ and $\RR_{++}=\{x\in \RR\mid x> 0\}$, and similarly define $\ZZ_+,\ZZ_{++}$ for the integer space. For a positive integer $n$, we denote $[n]$ the set $\{1,2,...,n\}$. For $x\in\RR^n$, denote $x^+=\max\{x,0\}$. 
For two real-valued functions $f,g$, we denote $f=O(g)$ if there exists $M>0$ such that $|f(x)|\leq Mg(x)$ for any $x$ of interest. Denote $f=\Omega(g)$ if $g=O(f)$.

The rest of the paper is organized as follows. In Section \ref{sec:LR}, we review the related literature on assortment optimization and bundle utility modeling, highlighting our contributions. In Section \ref{sec:model}, we describe the unconstrained assortment optimization problem under the two-category MVMNL model and show that this problem is strongly NP-hard. In Section \ref{sec:half_appro}, we propose a simple 0.5-approximation algorithm to this problem, and illustrate its limitations. In Section \ref{sec:approx_framework}, we describe our partition-and-optimize approximation framework based on an LP relaxation and illustrate how to use it to derive constant approximation ratios. In addition, we modify this framework to achieve an approximation ratio arbitrarily close to the integrality gap of the LP relaxation. In Section \ref{sec:ext}, we discuss the hardness results of several extensions to the two-category MVMNL model. In Section \ref{sec:numerical}, we perform numerical experiments to show the effectiveness of our proposed approximation framework.

\section{Literature review}\label{sec:LR}
In this section, we review the literature related to this work. Specifically, we first review the literature on modeling the utility of product bundles, mainly in the marketing literature. Then, we briefly review the literature on assortment optimization for \textit{single-choice} models. Lastly, we review a few attempts to solve assortment optimization under customers' multi-purchase behavior. 

The marketing and economics literature documented different types of interactions between products in a bundle to determine the bundle's utility in a MVMNL model. In these papers, customers choose the utility-maximizing bundle under a MVMNL model, an analogy to how customers choose the utility-maximizing product under an MNL model. \cite{Chung2003} use the comparability-based balance model to model the utility of bundles, where the attributes of products are divided into comparable attributes and non-comparable attributes. The comparable attribute of products in a bundle contributes to bundle utility in various ways apart from simple summation. \cite{ma2012modeling} study the multi-category choice behavior for households, where different types of product interactions are considered, including product complementarity, and their model is validated using scanner data from a large grocery store. The main takeaway from these studies is that there are various types of product interactions that determine the perceived utility of a bundle, whose structures are complicated and hard to be exploited for optimization purposes. For a more comprehensive literature review on how the bundle utility is modeled in the MVMNL model, refer to \cite{rao2018emerging}, and \cite{agarwal2015interdisciplinary}. The applications of the MVMNL model in the marketing literature include empirical analysis of market baskets (\cite{russell2000analysis}), and pricing decisions for competing retailers (\cite{richards2018retail}), etc. See \cite{bel2015multivariate} for a more detailed review of the literature on the applications of the MVMNL model. However, there are very few papers that study assortment optimization under these models, as we will review next.

Some variants of the MVMNL model are also justified by real-world datasets. We point the readers to two papers that did numerical validation on models very similar to our model. \cite{benson2018discrete} constructed a model where a customer chooses subsets of products of at most size $k$, based on a MVMNL model. In their paper, a set $H$ of bundles are allowed to have arbitrary utility, while all the other bundles have utility equal to the sum of utilities of individual products. The size of set $H$ is given exogenously, and they propose a method to optimize the composition of set $H$ in order to best fit data. They show with 6 real-world datasets that as the given size of $H$ increases, the representing power of the multi-purchase model increases, while their optimal set $H$ do not lead to any structure of bundle utilities. \cite{tulabandhula2020multi} conduct an extensive numerical study on the performance of their proposed Bundle-MVL models, especially with a bundle size of two. As we discussed earlier, the difference between their model and ours is that in their model, there is no clear specification of multiple disjoint product categories. They show with $10$ datasets from a variety of choice scenarios that even at a bundle size of two, the BundleMVL choice model outperforms the sparse model of \cite{benson2018discrete} and the MNL model in both the model fit and expected revenue. Also, they show that if the bundle size is increased from two to three, the optimized expected revenue does not change significantly. This provides strong justifications for our choice of focusing on two categories in this paper. In Section \ref{sec:ext} of the paper, we discuss an extension to three categories, and show that a constant-factor polynomial-time approximation is theoretically impossible. 

There has been a vast stream of literature considering assortment optimization in the past decade. 
Since the pioneering work of \cite{talluri2004revenue}’s in assortment optimization under the MNL model by \cite{mcfadden1973conditional}, researchers have considered assortment optimization under different variants of the MNL model. We refer the readers to \cite{qi2020data} and \cite{kok2015assortment} for a comprehensive literature review. 

The assortment optimization of the Paired Combinatorial Logit (PCL) models bears some similarity to ours in problem formulations, although its modeling assumptions are quite different from our model, as customers still purchase one item at a time (\cite{ghuge2022constrained}, \cite{zhang2020assortment}). We shall see in Remark \ref{remark:PCL} in Section \ref{sec:approx_framework} that the structure of the problem is different, and the methods used in these papers cannot be extended easily to our problem. In Section \ref{sec:ext} of this paper, we also show the inapproximability of our problem with cardinality constraint, while the assortment optimization under the PCL model with cardinality constraint has a constant-factor polynomial-time approximation.

Prior to our paper, there are very few works that consider assortment optimization under the MVMNL model or similar models.  \cite{tulabandhula2020multi} study a Bundle-MVL model, with one product category, and the customer can choose a limited number of products into a bundle. As we mentioned earlier, the authors conduct a thorough numerical study on the empirical performance of the model on various datasets and show that with bundle size at most $2$, the performance of the model significantly surpasses the existing benchmark, in terms of both the model fit and the expected revenue from the resulting assortment optimization problem. They show that increasing allowed bundle size from $2$ does not have significant improvement over the model performance. The authors provide algorithms to solve the problem exactly, in exponential time.  \cite{lyu2021assortment} study assortment optimization based on the MVMNL model where customers purchase at most one product from each product category and develop an FPTAS under stronger assumptions of product interaction. For the customer who purchases bundles, they assume interactions of products between products only depend on the category and is a constant number for a two-category setting. In our model, we admit arbitrary product interaction in the bundle by allowing general structure on the bundle's utility. \cite{Ghoniem2016} study assortment and price optimization under asymmetric cross-selling effects. The authors formulate a mixed-integer-non-linear programming problem to solve the assortment and pricing optimization problem. 

%Based on the numerical experiments from \cite{tulabandhula2020multi} that model with a bundle size of two can improve the model fit and expected revenue substantially whereas increasing bundle size does not have significant improvement and the preceding discussion, we argue that our focus on the two-category MVMNL model is both theoretically interesting and practically relevant.  

We note that there are some other assortment optimization problems considered in the optimization literature where customers can purchase multiple products at a time. For example, \cite{ke2022cross} consider a choice model where customers make sequential purchase decisions among a primary product category and a second product category. \cite{feldman2021assortment} consider a multi-choice model where customers choose all products that have utility above a certain threshold. \cite{zhang2021assortment} study assortment optimization where the customer can decide to purchase multiple units of different products and derives an 0.5-approximation algorithm for the problem. In their model, a bundle's utility is the summation of utilities provided by different products, where the utility provided by adding a unit of product decreases as the number of units of the same product in the bundle increases. However, it does not consider product interaction within a bundle.

%\textit{Multi-choice} models have been studied extensively in marketing and economics literature. These models are aimed at analyzing the multi-purchase behavior of customers. One of the most popular models in this stream of literature is the Multivariate MNL (MVMNL) model, where the \textit{single-choice} MNL model is generalized to \textit{multi-choice} scenarios. In this model, there are multiple product categories, and customers are allowed to choose at most one product from each category, to form a bundle. Each product bundle is assigned a separate utility parameter, which may or may not depend on the utility of the individual products inside the bundle. The customer still chooses the utility-maximizing product subsets, which could be an individual product or a bundle of products. 

%{\color{red} need to list the types of interactions between bundles documented in literature, and whether it increase/decrease bundle utility compared to summation?}

% \subsection*{Notation}
% $\RR_+$ is the set of nonnegative real numbers.

\section{Choice Model and Assortment optimization Problem}\label{sec:model}
This section presents the MVMNL model with two categories and its assortment optimization problem. Our model can be readily extended to an arbitrary number of categories.

Consider a retailer selling products from two categories: category 1 and category 2. Category 1 has $n$ products labeled $1,...,n$, and category 2 has $m$ products labeled $1,...,m$. Denote $ \mathbf{N}=\{1,...,n\}$ and $ \mathbf{M}=\{1,...,m\}$ the sets of products of category 1 and category 2, respectively. Denote $ \mathbf{N}_+ =\mathbf{N}\cup\{0\}$ and $ \mathbf{M}_+= \mathbf{M} \cup\{0\}$, where the index $0$ is used to indicate the no-purchase option. Let $p_i\in \RR_+$ be the unit price of product $i\in \mathbf{N}$, and $q_j\in \RR_+$ be the unit price of product $j\in \mathbf{M}$. Without loss of generality, we assume that $p_1\geq p_2\geq \cdots\geq p_n$ and $q_1\geq q_2\geq \cdots \geq q_m$. Let $u_{ij}\in \RR_+$ be the preference weight of products bundle $(i,j)\in \mathbf{N}\times \mathbf{M}$ and  $u_{i0},u_{0j}\in \RR_+$ be the preference weights of products $i\in \mathbf{N},j\in \mathbf{M}$ respectively. We view bundle $(i,0)$ (similarly for $(0,j)$) as the choice of a single item $i$ in the first category. For notation convenience, let $p_0=q_0=0$ and $u_{00}=1$. An assortment of products is a subset $A\times B\subseteq \mathbf{N}\times \mathbf{M}$ with $A\subseteq \mathbf{N}$ and $B\subseteq \mathbf{M}$. We represent an assortment $A\times B$ by  binary variables $x\in \{0,1\}^n,y\in\{0,1\}^m$ such that $x_i=1$ if $i\in A$ and $x_i=0$ if $i\in \mathbf{N}\setminus A$ ($y$ is defined similarly). Given an assortment $(x,y)$, we assume that the probability of purchasing bundle $(i,j)\in  \mathbf{N}_+ \times  \mathbf{M}_+ $ is given by
\begin{align}\label{eq:choice_prob}
\PP(i,j|x,y)=\frac{u_{ij}x_iy_j}{\sum_{i\in  \mathbf{N}_+ ,j\in  \mathbf{M}_+ }u_{ij}x_iy_j},   
\end{align}
where $x_0=y_0=1$. 
% Note that we allow customers to purchase any possible combinations of items given the assortment $(x,y)$, including only a single product. THIS IS CONFUSING
Note that we allow customers to purchase a single product only. Moreover, the model can capture various product interactions in a bundle. For example, incompatibility of bundle $(i,j)$ can be captured by setting $u_{ij}=0$. 

The choice probability of our model follows the widely-used MVMNL model in the literature (see the reference in Section \ref{sec:LR}). It can be derived from the assumption that the conditional probability of purchasing a product in one category follows the MNL model, given the purchasing decisions for another category. 
This assumption also appears in \cite{tulabandhula2020multi} and \cite{russell2000analysis}. Without loss of generality, we consider offered assortment $\mathbf{N}_+\times \mathbf{M}_+$. Assume that given a choice of product $i$ from category 1, the utility of purchasing bundle $(i,j)$ is 
$$U_{j|i}=a_{ij}+\epsilon_{j|i},\ j\in  \mathbf{M}_+ ,$$
where we assume that $\epsilon_{j|i},\ j\in  \mathbf{M}_+ $ follow \emph{i.i.d.}~Gumbel distributions with mean zero and scale parameter one. Similarly, we define $U_{i|j}=a_{ij}+\epsilon_{i|j},\ i\in  \mathbf{N}_+ $. Here, $a_{ij}$ can be regarded as the intrinsic value of bundle $(i,j)$. Assume $a_{00}=0$. It follows that the conditional probability of purchasing bundle $(i,j)$ given purchasing product $i$ from category 1 is
$$P_{j|i}=\frac{e^{a_{ij}}}{\sum_{k\in  \mathbf{M}_+ }e^{a_{ik}}}.$$
Similar for the conditional probability of purchasing bundle $(i,j)$ given purchasing product $j$ from category 2. We denote $P_{ij}$ the purchasing probability of bundle $(i,j)$ and $P^1_i=\sum_{j\in \mathbf{M}_+}P_{ij}$ the marginal probability that product $i$ in category $1$ is purchased (similar for $P^2_j$). We then have
\begin{align*}
    P_{ij}=P_{i|j}P^2_j,~~P_{0j}=P_{0|j}P^2_j,~~P_{0j}=P_{j|0}P^1_0,~~P_{00}=P_{0|0}P^1_0.
\end{align*}
The probability of purchasing bundle $(i,j)$ is
\begin{align*}
    P_{ij}=P_{i|j}P^2_j=P_{i|j}\frac{P_{0j}}{P_{0|j}}=\frac{P_{i|j}}{P_{0|j}}\frac{P_{j|0}}{P_{0|0}}P_{00}=\frac{e^{a_{ij}}}{e^{a_{0j}}}e^{a_{0j}}P_{00}=e^{a_{ij}}P_{00}.
\end{align*}
Since the sum of the probabilities $P_{ij}, i\in  \mathbf{N}_+ ,j\in  \mathbf{M}_+ $ equals one, we have $P_{00}=\frac{1}{\sum_{i\in  \mathbf{N}_+ ,j\in  \mathbf{M}_+ }e^{a_{ij}}}$. Therefore, $P_{ij}$ is given by equation \eqref{eq:choice_prob}. 

The objective of the retailer is to choose an assortment to maximize the expected revenue, i.e., 
\begin{align}\label{prob:IP}
\begin{aligned}
\max\ & \pi(x,y)\\
\text{s.t. } & x\in \{0,1\}^n,\ y\in \{0,1\}^m,
\end{aligned}
\end{align}
where 
$$\pi(x,y)=\sum_{i\in  \mathbf{N}_+ ,j\in  \mathbf{M}_+ }\PP(i,j|x,y)(p_i+q_j)=\frac{\sum_{i\in  \mathbf{N}_+ ,j\in  \mathbf{M}_+ }u_{ij}x_iy_j(p_i+q_j)}{\sum_{i\in  \mathbf{N}_+ ,j\in  \mathbf{M}_+ }u_{ij}x_iy_j}.$$

In problem \eqref{prob:IP}, we set the price of bundles to be the summation of the prices of products in it. We focus on this pricing scheme in the majority of the paper, and discuss general bundle prices in Section \ref{subsec:gap-e} and Section \ref{subsec:general price}.
The following result shows that problem \eqref{prob:IP} is strongly NP-hard. This indicates that this problem is harder than some closely related assortment optimization problems studied in the literature (e.g., \citealt{lyu2021assortment}), where FPTAS exists. 
% Unless P=NP, no PTAS exists for this problem.
\begin{theorem}\label{thm:NPhard}
Problem (\ref{prob:IP}) is strongly NP-hard.
\end{theorem}

The proof is provided in Appendix \ref{Appendix:proof}. It constructs a reduction from the \textsc{Max-DiCut} problem on directed acyclic graphs, which is known to be strongly NP-hard (\citealt{lampis2011algorithmica}). Moreover, this result rules out the existence of FPTAS for problem \eqref{prob:IP}.
Motivated by Theorem \ref{thm:NPhard}, we focus on the design of approximation algorithms in this paper.

\section{An 0.5-Approximation Algorithm}\label{sec:half_appro}
% In this section, we provide a simple algorithm with a 0.5-approximation guarantee. This algorithm is based on the closed form solution of a special case of problem \eqref{prob:IP}, which is presented in the following lemma. 

In this section, we first show that a special case of the assortment optimization problem has a closed-form optimal solution which, we term \textit{adjusted-revenue-ordered} assortments. We then show that \textit{adjusted-revenue-ordered} assortments provides a 0.5-approximation of the original problem.

\begin{lemma}\label{lemma:q=0}
If $q_j=0$ for any $j\in \mathbf{M}$, then $(x,y)$ with 
\begin{align*}
\begin{cases}
x_i=1 & \text{if and only if}\quad p_i\geq \pi^p,\\
y_j=1 & \text{if and only if}\quad  \frac{\sum_{i\in \mathbf{N}}u_{ij}p_ix_i}{\sum_{i\in  \mathbf{N}_+ }u_{ij}x_i}\geq \pi^p,
\end{cases}
\end{align*}
is an optimal solution of problem \eqref{prob:IP}, where $\pi^p$ is the optimal objective value of problem \eqref{prob:IP} with $q_j=0,j\in \mathbf{M}$.
\end{lemma}
\proof{Proof of Lemma \ref{lemma:q=0}.}
It is easy to see that a solution $(x,y)$ is optimal in problem \eqref{prob:IP} if and only if it is an optimal solution of the following problem
\begin{align}\label{eq:q=0}
\max_{x\in \{0,1\}^n,y\in \{0,1\}^m} \sum_{i\in  \mathbf{N}_+ }(p_i-\pi^p)x_i\sum_{j\in  \mathbf{M}_+ } u_{ij}y_j.   
\end{align}
For any given $y\in \{0,1\}^m$, it is optimal to have $x_i=1$ if $p_i\geq \pi^p$ and $x_i=0$ if $p_i<\pi^p$ in problem \eqref{eq:q=0}. Fixing such value of $x$, the objective in \eqref{eq:q=0} can be written as $\sum_{j\in  \mathbf{M}_+ }y_j\sum_{i\in  \mathbf{N}_+ }u_{ij}(p_i-\pi^p)x_i$, which is maximized for $y_j=1$ with $j$ satisfying  $\sum_{i\in  \mathbf{N}_+ }u_{ij}(p_i-\pi^p)x_i\geq 0$ or equivalently $\frac{\sum_{i\in \mathbf{N}}u_{ij}p_ix_i}{\sum_{i\in  \mathbf{N}_+ }u_{ij}x_i}\geq \pi^p$.
\Halmos
\endproof

Lemma \ref{lemma:q=0} shows that for the special case of $q_j=0,j\in \mathbf{M}$, one of the solutions $(x^{kl},y^{kl}),k=1,...,n,\ l=1,...,m$ is optimal, where
\begin{align}\label{eq:adjust_revenue_order}
\begin{aligned}
\begin{cases}
x_i^{kl}=1 & \text{if and only if}\quad i\leq k,\\
y_j^{kl}=1 & \text{if and only if} \quad \frac{\sum_{i\in \mathbf{N}}u_{ij}p_ix_i^{kl}}{\sum_{i\in  \mathbf{N}_+ }u_{ij}x_i^{kl}} \text{ is among the largest $l$ of } \left\{\frac{\sum_{i\in \mathbf{N}}u_{is}p_ix_i^{kl}}{\sum_{i\in  \mathbf{N}_+ }u_{is}x_i^{kl}}\bigg | s\in \mathbf{M} \right\}.
\end{cases}
\end{aligned}
\end{align}

Similar result holds for the case of $p_i=0,i\in \mathbf{N}$.
The assortments in \eqref{eq:adjust_revenue_order} (and those for the case of $p_i=0,i\in \mathbf{N}$) are referred to as \textit{adjusted-revenue-ordered} assortments. Similar to the well-known \textit{revenue-ordered} assortments (e.g., \citealt{talluri2004revenue}), an adjusted-revenue-ordered assortment selects a revenue-ordered assortment from one category but selects a level set of products from the other category based on the order of certain weighted revenues.  

We now present the 0.5-approximation algorithm by utilizing the adjusted-revenue-ordered assortments in Algorithm \ref{alg:half} . The idea of the algorithm is to solve two special cases of the problem \eqref{prob:IP} by setting either all $p_i$'s or all $q_j$'s to be $0$. According to Lemma \ref{lemma:q=0}, these two special cases can be solved efficiently by comparing the revenue of all the adjusted-revenue-ordered assortments. The algorithm outputs one of the adjusted-revenue-ordered assortments obtained from the two special cases that has the largest revenue. 

\begin{algorithm}
\SetKwInOut{Input}{Input}
\Input{$u_{ij},p_i,q_j$ for $i\in  \mathbf{N}_+ ,j\in  \mathbf{M}_+ $}
\BlankLine
Solve problem \eqref{prob:IP} for the case of $q_j=0,j\in \mathbf{M}$ and obtain the optimal objective value $\pi^p$\\
Solve problem \eqref{prob:IP} for the case of $p_i=0,i\in \mathbf{N}$ and obtain the optimal objective value $\pi^q$\\
\If{$\pi^p\geq \pi^q$}{output the optimal solution of the case $q_j=0,j\in \mathbf{M}$} \Else{output the optimal solution of the case $p_i=0,i\in \mathbf{N}$}
\caption{}\label{alg:half}
\end{algorithm}

The following result shows that Algorithm \ref{alg:half} is a 0.5-approximation algorithm.

\begin{theorem}\label{thm:half}
Let $(x^{\ARO},y^{\ARO})$ be the solution obtained from Algorithm \ref{alg:half}, then $\pi(x^{\ARO},y^{\ARO})\geq 0.5\pi^*$, where $\pi^*$ is the optimal objective value of problem \eqref{prob:IP}.
\end{theorem}
\proof{Proof of Theorem \ref{thm:half}.}
It is clear that
\begin{align*}
    \pi^*= &\max_{x,y}  \frac{\sum_{i\in  \mathbf{N}_+ ,j\in  \mathbf{M}_+ } u_{ij}x_iy_j(p_i+q_j)}{\sum_{i\in  \mathbf{N}_+ ,j\in  \mathbf{M}_+ } u_{ij}x_iy_j}\\ 
    \leq & \max_{x,y}  \frac{\sum_{i\in  \mathbf{N}_+ ,j\in  \mathbf{M}_+ } u_{ij}p_ix_iy_j}{\sum_{i\in  \mathbf{N}_+ ,j\in  \mathbf{M}_+ } u_{ij}x_iy_j}
    +  \max_{x,y}  \frac{\sum_{i\in  \mathbf{N}_+ ,j\in  \mathbf{M}_+ } u_{ij}q_jx_iy_j}{\sum_{i\in  \mathbf{N}_+ ,j\in  \mathbf{M}_+ } u_{ij}x_iy_j}\\ 
    =& \pi^p+\pi^q\\
    \leq &  2\max\{\pi^p,\pi^q\}.
\end{align*}
Without loss of generality, assume $\max\{\pi^p,\pi^q\}=\pi^p$. We then have 
$$\pi(x^{\ARO},y^{\ARO})=\frac{\sum_{i\in  \mathbf{N}_+ ,j\in  \mathbf{M}_+ } u_{ij}x_i^{\ARO}y_j^{\ARO}(p_i+q_j)}{\sum_{i\in  \mathbf{N}_+ ,j\in  \mathbf{M}_+ } u_{ij}x_i^{\ARO}y_j^{\ARO}}\geq  \frac{\sum_{i\in  \mathbf{N}_+ ,j\in  \mathbf{M}_+ } u_{ij}p_ix_i^{\ARO}y_j^{\ARO}}{\sum_{i\in  \mathbf{N}_+ ,j\in  \mathbf{M}_+ } u_{ij}x_i^{\ARO}y_j^{\ARO}} =\pi^p.$$ 
Hence, $\pi(x^{\ARO},y^{\ARO})\geq 0.5\pi^*$.
\Halmos
\endproof

\begin{remark}
From the proof of Theorem \ref{thm:half}, the instance-dependent approximation ratio of Algorithm \ref{alg:half} is at least $\frac{\max\{\pi^p,\pi^q\}}{\pi^p+\pi^q}$. This ratio is large when the prices of one category is much larger than the prices of the other category.
\end{remark}

\subsection*{Limitations of Algorithm 1}

Algorithm \ref{alg:half}, though simple, can only guarantee half of the optimal revenue in the worst case.
\begin{example}
Let $\varepsilon,M$ be positive numbers with $M\varepsilon=2$. Consider a problem instance with $n=3$, $m=3$, $p_1=q_1=1+\varepsilon,\ p_2=q_2=1,\ p_3=q_3=0$, $u_{00}=1,\ u_{22}=u_{13}=u_{31}=M,\ u_{ij}=0 \text{ otherwise}$.
\end{example}

For this problem instance, one can verify that when $M\rightarrow \infty$, the optimal assortment for $\pi^p$ is $x_1=\{1,0,0\},y_1=\{0,0,1\}$ with $\pi^p=\frac{M(1+\varepsilon)}{M+1}$. The optimal assortment obtained by solving for $\pi^q$ is $x_2=\{0,0,1\},y_2=\{1,0,0\}$ with $\pi^q=\frac{M(1+\varepsilon)}{M+1}$. The optimal assortment to the original problem is $x^*=\{0,1,0\},y^*=\{0,1,0\}$ with an objective value of $\pi^*=\frac{2M}{M+1}$. We can see that $\pi(x_1,y_1)=\pi(x_2,y_2)=\frac{M(1+\varepsilon)}{M+1}$ and $\max\{\pi(x_1,y_1),\pi(x_2,y_2)\}/\pi^*\rightarrow 1/2$ when $M\rightarrow \infty$.

\section{An Approximation Framework Based on LP Relaxations}\label{sec:approx_framework}
In this section, we propose an approximation framework based on an LP relaxation of problem \eqref{prob:IP}, termed \textit{partition-and-optimize}. This framework allows us to obtain approximation algorithms with approximation ratios close to the integrality gap of the LP relaxation. We first present the LP relaxation for the assortment optimization problem and show that the integrality gap of the LP relaxation is at most 0.75. Then, in Section \ref{subsec:approx framework}, we formally present the partition-and-optimize approximation framework that constructs $K$ different assortments based on the LP optimal solution. In Section \ref{subsec:K46}, we apply our approximation framework to $K=4,6$ and show that we can get an approximation guarantee that almost matches the integrality gap.
% Finally, in Section \ref{subsec:gap-e}, we show how to extend our approximation framework to arbitrarily close to the integrality gap.
Finally, in Section \ref{subsec:gap-e}, we propose an approximation scheme by modifying the partition-and-optimize framework to approximate the problem with a factor arbitrarily closed to the integrality gap.

An LP relaxation of problem \eqref{prob:IP} can be obtained as follows. Let $z_{ij}=x_iy_j\in \{0,1\}$ for $(i,j)\in \mathbf{N}\times \mathbf{M}$ and $w=(1+\sum_{i\in \mathbf{N},j\in \mathbf{M}}u_{ij}z_{ij}+\sum_{i\in \mathbf{N}}u_{i0}x_i+\sum_{j\in \mathbf{M}}u_{0j}y_j)^{-1}$. Note that $\{(x_i,y_j,z_{ij}) \mid z_{ij}=x_iy_j;x_i,y_j\in \{0,1\}\}$ can be relaxed to $\{(x_i,y_j,z_{ij}) \mid 0\leq z_{ij}\leq x_i, 0\leq z_{ij}\leq y_j, z_{ij}\geq x_i+y_j-1\}$ (e.g., \citealt{padberg1989boolean}). By scaling $x_i,y_j,z_{ij}$ with $w$, problem \eqref{prob:IP} has the following LP relaxation:
\begin{subequations}\label{prob:LP}
\begin{align}
\max_{x,y,z,w} \ &\sum_{i\in \mathbf{N},j\in \mathbf{M}} u_{ij}(p_i+q_j)z_{ij}+\sum_{i\in \mathbf{N}}u_{i0}p_ix_i+\sum_{j\in \mathbf{M}}u_{0j}q_jy_j \\
\text{s.t. } & w+\sum_{i\in \mathbf{N},j\in \mathbf{M}} u_{ij}z_{ij}+\sum_{i\in \mathbf{N}}u_{i0}x_i+\sum_{j\in \mathbf{M}}u_{0j}y_j=1 \label{prob:LP_cons1}\\
 & z_{ij}\leq x_i,\ i\in \mathbf{N},\ j\in \mathbf{M} \label{prob:LP_cons2}\\
 & z_{ij}\leq y_j,\ i\in \mathbf{N},\ j\in \mathbf{M} \label{prob:LP_cons3}\\
& z_{ij} \geq x_i+y_j-w,\ i\in \mathbf{N},\ j\in \mathbf{M} \label{prob:LP_cons4}.
\end{align}
\end{subequations}

For an instance $\mathcal{I}$ of problem \eqref{prob:IP}, denote the optimal objective values of problems \eqref{prob:IP} and \eqref{prob:LP} by $\pi^*_{\mathcal{I}}$ and $r^*_{\mathcal{I}}$, respectively, and define the optimal solution of LP relaxation \eqref{prob:LP} as $(w^{\mathrm{LP}},x^{\mathrm{LP}},y^{\mathrm{LP}},z^{\mathrm{LP}})$. We may suppress the subscript $\mathcal{I}$ when the instance is clear from the context. The integrality gap of problem \eqref{prob:LP} for an instance $\mathcal{I}$ is defined by the ratio $\Gap_{\mathcal{I}}=\frac{\pi^*_{\mathcal{I}}}{r^*_{\mathcal{I}}}\in [0,1]$. The integrality gap of problem \eqref{prob:LP} is defined as the infimum of $\Gap_{\mathcal{I}}$ over all instances of the problem and is denoted by $\Gap$.
The following lemma characterizes the structures of the optimal solution of problem \eqref{prob:LP}, which are useful in our approximation framework. %As we observe in this lemma, the structure of the optimal solution to this LP is different compared to the PCL model. In the LP relaxation of \cite{zhang2020assortment}, the upper bounds on the cross term, counterparts of \eqref{prob:LP_cons2} and \eqref{prob:LP_cons3} are redundant at the LP's optimal solution.

\begin{lemma}\label{lemma:half-zij}
Let $(x,y,z,w)$ be an optimal basic feasible solution of problem \eqref{prob:LP}.
\begin{enumerate}[label=(\alph*)]
    \item It holds that $x_i,y_j,z_{ij} \in \{0,\frac{w}{2},w\}$ for all $i\in \mathbf{N},j\in \mathbf{M}$.
    \item  If $x_i=y_j=\frac{1}{2}w$ for some $i\in \mathbf{N},j\in \mathbf{M}$, then 
\begin{align*}
z_{ij}= \begin{cases}
    \frac{1}{2}w & \text{ if  } \quad p_i+q_j> r^*,\\
    0 & \text{ if  } \quad p_i+q_j< r^*,\\
    0 \text{ or }  \frac{1}{2}w & \text{ if  } \quad p_i+q_j= r^*.
    \end{cases}
\end{align*}
\end{enumerate}
\end{lemma}

\proof{Proof of Lemma \ref{lemma:half-zij}.} 
 (a) It is clear that $(\frac{x}{w},\frac{y}{w},\frac{z}{w})$ is a basic feasible solution of $\{0\leq z_{ij}\leq x_i,\ 0\leq z_{ij}\leq y_j,\ z_{ij}\geq x_i+y_j-1,\ i\in N,\ j\in M\}$ whose vertex can be proven to be $\{0,\frac{1}{2},1\}$-valued (follows exactly the same proof of Theorem 7 in \cite{padberg1989boolean}). 

(b) We only prove for the case $p_i+q_j<r^*$ as the other cases follow similar arguments. Suppose $z_{ij}=\frac{1}{2}w$. Changing the value of $z_{ij}$ to zero and increasing the value of $w$ and all the other $x_i,y_j,z_{ij}$'s at the same scale until the equality constraint \eqref{prob:LP_cons1} is satisfied increase the objective value of the LP, a contradiction.
\Halmos
\endproof

A basic feasible solution of problem \eqref{prob:LP} with some variable equals $\frac{w}{2}$ is referred to as a \textit{half-integral solution} or \textit{non-integral solution}; otherwise it is referred to as an \textit{integral solution}.
\begin{remark}\label{remark:PCL}
We would like to mention that an LP relaxation similar to \eqref{prob:LP} is employed in \cite{zhang2020assortment} to design approximation algorithms for the assortment optimization problem under the PCL model. It is tempting to follow the random rounding method of \cite{zhang2020assortment} to round the optimal solution of our LP relaxation, i.e., round $x_i,y_j=\frac{1}{2}w$ to $0$ and $w$ with equal probabilities. However, this will allow a $\frac{1}{4}$ probability to choose the ``bad" bundle $(i,j)$ with $p_i+q_j<r^*$ by Lemma \ref{lemma:half-zij}(b), which can heavily decrease the total revenue. Instead, this issue does not exist in \cite{zhang2020assortment} due to the special structure of the optimal LP solution.

% We would like to compare our LP relaxation with the LP relaxation for the assortment optimization problem under the PCL model in \cite{zhang2020assortment}. In the PCL model, an assortment is denoted by a set of binary variables $x_i\in\{0,1\},i\in N$. The customers first decide to make a purchase among one of the product nests $(i,j)\in N^2$, then make a choice among products in the chosen nest. The decision variable $y_{ij}\in\{0,1\}$ is defined with the interpretation that $y_{ij}=x_ix_j$. A McCormick relaxation is used to construct the LP relaxation for the assortment optimization problem in the PCL model. The LP relaxations look similar. However, as we observe in lemma \ref{lemma:half-zij}, the structure of the optimal solution to the LP relaxation is different. In \cite{zhang2020assortment}, the optimal cross-term $y_{ij}=(x_i+x_j-1)^+$ in the LP relaxation, due to the coefficient values for $y_{ij}$ in the objective function. On the other hand, the value of optimal $z_{ij}$ in \eqref{prob:LP} depends on $p_i+q_j$. So, randomly rounding the LP optimal solution following methods used for PCL model does not work for our problem.
\end{remark}

\begin{theorem}\label{thm:int_gap}
The integrality gap of problem \eqref{prob:LP} is at most 0.75.
\end{theorem}
\proof{Proof of Theorem \ref{thm:int_gap}.}
It suffices to provide an instance with an integrality gap at most 0.75. Let $M>0$ be a large number. Consider an instance with $n=m=4$, $p_1=q_1=\frac{3M}{4}$, $p_2=q_2=\frac{3}{4}$, $p_3=q_3=\frac{3}{8}$, $p_4=q_4=0$, $u_{14}=u_{41}=M^{-1}$, $u_{23}=u_{32}=2$, and $u_{24}=u_{33}=u_{34}=u_{42}=u_{43}=u_{44}=M$. All the other $u_{ij}$'s are set to be zero. Figure \ref{fig:int-gap} shows the values of the parameters. In this figure, the $i$-th row (column) represents the product $i$ in category 1 (category 2) for $i=0,1,...,4$, where product 0 represents the no-purchasing option.
\begin{figure}[htbp]
\centering

\begin{tikzpicture}[x=0.75pt,y=0.75pt,yscale=-1,xscale=1]
%uncomment if require: \path (0,667); %set diagram left start at 0, and has height of 667

%Straight Lines [id:da974780124927954] 
\draw    (83,160) -- (143,160) ;
%Straight Lines [id:da3211810243026667] 
\draw    (143,160) -- (203,160) ;
%Straight Lines [id:da33564528568929797] 
\draw    (203,160) -- (263,160) ;
%Straight Lines [id:da5043493358375941] 
\draw    (263,160) -- (383,160) ;
%Straight Lines [id:da001457459398999994] 
\draw    (83,160) -- (83,220) ;
%Straight Lines [id:da09313037013253345] 
\draw    (83,220) -- (83,280) ;
%Straight Lines [id:da3662329980234662] 
\draw    (83,280) -- (83,340) ;
%Straight Lines [id:da581233160266938] 
\draw    (83,340) -- (83,460) ;
%Straight Lines [id:da19219044747000047] 
\draw    (323,160) -- (323,460) ;
%Straight Lines [id:da5260961311983376] 
\draw    (83,400) -- (383,400) ;
%Straight Lines [id:da8388973770453783] 
\draw    (263,160) -- (263,460) ;
%Straight Lines [id:da023968265825428325] 
\draw    (203,160) -- (203,460) ;
%Straight Lines [id:da15617738748698073] 
\draw    (143,160) -- (143,460) ;
%Straight Lines [id:da26938798614493287] 
\draw    (83,220) -- (383,220) ;
%Straight Lines [id:da607307849147029] 
\draw    (83,280) -- (383,280) ;
%Straight Lines [id:da9484399587922521] 
\draw    (83,340) -- (383,340) ;
%Straight Lines [id:da2116922672173256] 
\draw    (383,160) -- (383,460) ;
%Straight Lines [id:da8924466198051733] 
\draw    (83,460) -- (383,460) ;

% Text Node
\draw (108,183) node [anchor=north west][inner sep=0.75pt]    {$1$};
% Text Node
\draw (11,242) node [anchor=north west][inner sep=0.75pt]  [font=\footnotesize]  {$p_{1} =3M/4$};
% Text Node
\draw (24,302) node [anchor=north west][inner sep=0.75pt]  [font=\footnotesize]  {$p_{2} =3/4$};
% Text Node
\draw (22,362) node [anchor=north west][inner sep=0.75pt]  [font=\footnotesize]  {$p_{3} =3/8$};
% Text Node
\draw (35,419) node [anchor=north west][inner sep=0.75pt]  [font=\footnotesize]  {$p_{4} =0$};
% Text Node
\draw (332,132) node [anchor=north west][inner sep=0.75pt]  [font=\footnotesize]  {$q_{4} =0$};
% Text Node
\draw (265,129) node [anchor=north west][inner sep=0.75pt]  [font=\footnotesize]  {$q_{3} =3/8$};
% Text Node
\draw (205,129) node [anchor=north west][inner sep=0.75pt]  [font=\footnotesize]  {$q_{2} =3/4$};
% Text Node
\draw (138,129) node [anchor=north west][inner sep=0.75pt]  [font=\footnotesize]  {$q_{1} =3M/4$};
% Text Node
\draw (90,129) node [anchor=north west][inner sep=0.75pt]  [font=\footnotesize]  {$q_{0}=0$};
% Text Node
\draw (35,183) node [anchor=north west][inner sep=0.75pt]  [font=\footnotesize]  {$p_{0}=0$};
% Text Node
\draw (157,419) node [anchor=north west][inner sep=0.75pt]    {$M^{-1}$};
% Text Node
\draw (228,362) node [anchor=north west][inner sep=0.75pt]    {$2$};
% Text Node
\draw (335,239) node [anchor=north west][inner sep=0.75pt]    {$M^{-1}$};
% Text Node
\draw (288,303) node [anchor=north west][inner sep=0.75pt]    {$2$};
% Text Node
\draw (222,421) node [anchor=north west][inner sep=0.75pt]    {$M$};
% Text Node
\draw (283,421) node [anchor=north west][inner sep=0.75pt]    {$M$};
% Text Node
\draw (342,420) node [anchor=north west][inner sep=0.75pt]    {$M$};
% Text Node
\draw (283,362) node [anchor=north west][inner sep=0.75pt]    {$M$};
% Text Node
\draw (343,303) node [anchor=north west][inner sep=0.75pt]    {$M$};
% Text Node
\draw (343,362) node [anchor=north west][inner sep=0.75pt]    {$M$};

\end{tikzpicture}
\caption{Integrality gap instance}
\label{fig:int-gap}
\end{figure}

Consider a feasible solution of the LP \eqref{prob:LP} as follows. $x_1=x_2=x_3=x_4= y_1=y_2=y_3=y_4= z_{14}=z_{23}=z_{32}=z_{41}=\frac{w}{2}$, all the other $z_{ij}=0$, and $w$ is chosen so that the equality constraint \eqref{prob:LP_cons1} holds. The LP objective value for this solution is
\begin{align*}
    \frac{0.5\times(0.75M\times M^{-1}+(3/4+3/8)\times 2+(3/4+3/8)\times 2+0.75M\times M^{-1})}{1+0.5\times (2+2+M^{-1}+M^{-1})}=\frac{3}{3+M^{-1}},
\end{align*}
which implies $r^*\geq \frac{3}{3+M^{-1}}$.
Next, we prove that when $M$ goes to infinity, $\pi^*$ goes to $\frac{3}{4}$, i.e. the LP objective of the best integral solutions goes to $\frac{3}{4}$. We note that for all integral solutions, $\frac{z_{ij}}{w}=\frac{x_i}{w}\frac{y_j}{w}$. If $x_i,y_j$'s are selected such that at least two of $z_{14},z_{23},z_{32},z_{41}$ equal $w$, then there exists $z_{ij}=w$ for some $u_{ij}=M$. Since $p_i+q_j$ for these bundles $(i,j)$ with $u_{ij}=M$ are less than or equal to $\frac{3}{4}$, as $M$ goes to infinity, the object value goes to at most $\frac{3}{4}$, and one such integral solution is $x=\{0,w,w,0\},y=\{0,0,w,0\}$. If $x_i,y_j$'s are selected such that at most one of $z_{14},z_{23},z_{32},z_{41}$ equals $w$, then one can verify that the LP objective is at most $\frac{3}{4}$. Therefore, as $M$ goes to infinity, $\frac{\pi^*}{r^*}\leq \frac{3}{4}$.
\Halmos
\endproof

Theorem \ref{thm:int_gap} shows that in the worst case, no rounding algorithm based on the LP relaxation \eqref{prob:LP} can achieve an approximation ratio larger than 0.75. Interestingly, the approximation framework presented in the next subsection can be used to obtain a rounding algorithm with an approximation ratio of 0.74, which almost matches the integrality gap. 
% So, our approximation framework almost closed the integrality gap for the LP relaxation.  

\subsection{Partition-and-Optimize Approximation Framework}\label{subsec:approx framework}

This subsection presents the \textit{partition-and-optimize} approximation framework which provides a systematic way to round half-integral optimal solutions of problem \eqref{prob:LP}.

The framework can be divided into three steps. 
First, solve the LP relaxation \eqref{prob:LP} and obtain an optimal solution and the optimal objective $r^*$. If the optimal solution is non-integral, we \textit{partition} products with half-integral variables into $K$ level sets for both categories, defined by some cutoff points $b_1,\dots,b_K$, and construct $K$ different integral solutions using combinations of different level sets for the two categories.
Second, we use a quadratic  program to compute the approximation ratio of the $K$ integral solutions given the cutoff points of the level sets.  
Third, the cutoff points of the level sets are \textit{optimized} to get the highest approximation ratio.

Let $K$ be a positive integer and $(\Tilde{x}^{\mathrm{LP}},\Tilde{y}^{\mathrm{LP}},\Tilde{z}^{\mathrm{LP}})$ be a non-integral optimal solution $(x^{\mathrm{LP}},y^{\mathrm{LP}},z^{\mathrm{LP}})$ of problem \eqref{prob:LP} divided by $w ^{\LP}$. Note that $(\Tilde{x}^{\mathrm{LP}},\Tilde{y}^{\mathrm{LP}},\Tilde{z}^{\mathrm{LP}})$ is $\{0,\frac{1}{2},1\}$-valued by Lemma \ref{lemma:half-zij}. We now formally present the framework.

\subsubsection*{Step I: rounding.}

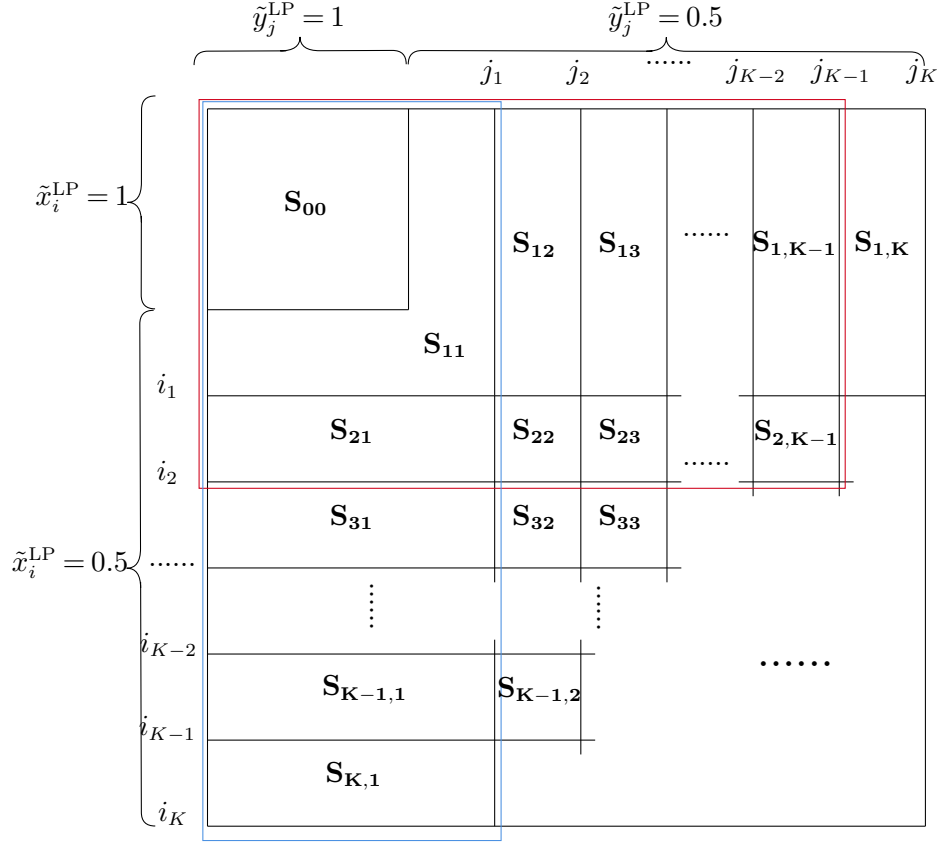
\begin{figure}[htbp]
\centering
\begin{tikzpicture}[x=0.75pt,y=0.75pt,yscale=-1,xscale=1]
%uncomment if require: \path (0,667); %set diagram left start at 0, and has height of 667

%Straight Lines [id:da7553561894593463] 
\draw    (133.93,90.69) -- (496,90.69) ;
%Straight Lines [id:da9484011462358262] 
\draw    (133.93,90.69) -- (133.93,452.76) ;
%Straight Lines [id:da7086821703172028] 
\draw    (235.31,90.69) ;
%Straight Lines [id:da31417570218742163] 
\draw    (235.31,90.69) -- (235.31,192.07) ;
%Straight Lines [id:da2623896977643876] 
\draw    (278.76,90.69) -- (278.76,235.52) ;
%Straight Lines [id:da01626958952376878] 
\draw    (322.21,90.69) -- (322.21,329.66) ;
%Straight Lines [id:da4400989204055461] 
\draw    (133.93,192.07) -- (235.31,192.07) ;
%Straight Lines [id:da5781242922591792] 
\draw    (133.93,235.52) -- (278.76,235.52) ;
%Straight Lines [id:da7075257194822628] 
\draw    (133.93,278.97) -- (372.9,278.97) ;
%Shape: Brace [id:dp750734383833507] 
\draw  [color={rgb, 255:red, 0; green, 0; blue, 0 }  ,draw opacity=1 ] (495.42,68.97) .. controls (495.43,64.3) and (493.11,61.96) .. (488.44,61.95) -- (375.31,61.63) .. controls (368.64,61.61) and (365.32,59.27) .. (365.33,54.6) .. controls (365.32,59.27) and (361.98,61.59) .. (355.31,61.57)(358.31,61.58) -- (242.18,61.26) .. controls (237.52,61.25) and (235.18,63.57) .. (235.17,68.24) ;
%Shape: Brace [id:dp4742947582128445] 
\draw  [color={rgb, 255:red, 0; green, 0; blue, 0 }  ,draw opacity=1 ] (234.15,68.17) .. controls (234.13,63.5) and (231.79,61.18) .. (227.12,61.2) -- (190.63,61.34) .. controls (183.96,61.37) and (180.62,59.05) .. (180.6,54.38) .. controls (180.62,59.05) and (177.3,61.39) .. (170.63,61.42)(173.63,61.41) -- (134.13,61.57) .. controls (129.46,61.59) and (127.14,63.93) .. (127.16,68.6) ;
%Shape: Brace [id:dp11453161645264887] 
\draw  [color={rgb, 255:red, 0; green, 0; blue, 0 }  ,draw opacity=1 ] (107.68,84.17) .. controls (103.01,84.16) and (100.68,86.49) .. (100.67,91.16) -- (100.6,127.89) .. controls (100.59,134.56) and (98.25,137.89) .. (93.58,137.88) .. controls (98.25,137.89) and (100.58,141.22) .. (100.57,147.89)(100.57,144.89) -- (100.5,184.62) .. controls (100.49,189.29) and (102.82,191.62) .. (107.49,191.63) ;
%Shape: Brace [id:dp08415724843102645] 
\draw  [color={rgb, 255:red, 0; green, 0; blue, 0 }  ,draw opacity=1 ] (106.77,192.36) .. controls (102.1,192.37) and (99.78,194.71) .. (99.79,199.38) -- (100.15,312.53) .. controls (100.17,319.2) and (97.85,322.54) .. (93.18,322.55) .. controls (97.85,322.54) and (100.19,325.86) .. (100.21,332.53)(100.2,329.53) -- (100.57,445.68) .. controls (100.58,450.35) and (102.92,452.67) .. (107.59,452.66) ;
%Straight Lines [id:da05530779941295827] 
\draw    (278.76,235.52) -- (278.76,250) -- (278.76,329.66) ;
%Straight Lines [id:da2870680674364816] 
\draw    (278.76,235.52) -- (372.9,235.52) ;
%Straight Lines [id:da035427399338794574] 
\draw    (365.66,90.69) -- (365.66,235.52) ;
%Straight Lines [id:da8052448614902712] 
\draw    (133.93,322.41) -- (372.9,322.41) ;
%Straight Lines [id:da8782546206990061] 
\draw    (365.66,235.52) -- (365.66,329.66) ;
%Straight Lines [id:da7211610967912405] 
\draw    (409.1,90.69) -- (409.1,286.21) ;
%Straight Lines [id:da1241984471120241] 
\draw    (452.55,90.69) -- (452.55,286.21) ;
%Straight Lines [id:da8658477374330951] 
\draw    (133.93,365.86) -- (329.45,365.86) ;
%Straight Lines [id:da5081963157680915] 
\draw    (133.93,409.31) -- (329.45,409.31) ;
%Straight Lines [id:da9615810557801288] 
\draw    (401.86,278.97) -- (459.79,278.97) ;
%Straight Lines [id:da017587403925572787] 
\draw    (278.76,358.62) -- (278.76,452.76) ;
%Straight Lines [id:da5679978404328951] 
\draw    (496,90.69) -- (496,452.76) ;
%Straight Lines [id:da3569466137729853] 
\draw    (322.21,358.62) -- (322.21,416.55) ;
%Straight Lines [id:da08509676250521703] 
\draw    (133.93,452.76) -- (496,452.76) ;
%Straight Lines [id:da3087998753868717] 
\draw    (401.86,235.52) -- (496,235.52) ;
%Shape: Rectangle [id:dp7055357306321268] 
\draw  [color={rgb, 255:red, 208; green, 2; blue, 27 }  ,draw opacity=1 ] (129.69,86) -- (455.55,86) -- (455.55,282.21) -- (129.69,282.21) -- cycle ;
%Shape: Rectangle [id:dp0001988142476951893] 
\draw  [color={rgb, 255:red, 74; green, 144; blue, 226 }  ,draw opacity=1 ] (131.59,87.07) -- (282,87.07) -- (282,460) -- (131.59,460) -- cycle ;

% Text Node
\draw (372.24,151.97) node [anchor=north west][inner sep=0.75pt]   [align=left] {......};
% Text Node
\draw (218.86,328.33) node [anchor=north west][inner sep=0.75pt]  [rotate=-90] [align=left] {......};
% Text Node
\draw (409.83,367.97) node [anchor=north west][inner sep=0.75pt]  [font=\Large] [align=left] {......};
% Text Node
\draw (372.24,267.83) node [anchor=north west][inner sep=0.75pt]   [align=left] {......};
% Text Node
\draw (102.86,318.52) node [anchor=north west][inner sep=0.75pt]   [align=left] {......};
% Text Node
\draw (353.41,65.07) node [anchor=north west][inner sep=0.75pt]   [align=left] {......};
% Text Node
\draw (333.28,329.05) node [anchor=north west][inner sep=0.75pt]  [rotate=-90] [align=left] {......};
% Text Node
\draw (171.21,130.69) node [anchor=north west][inner sep=0.75pt]    {$\mathbf{S_{00}}$};
% Text Node
\draw (241.45,202.38) node [anchor=north west][inner sep=0.75pt]    {$\mathbf{S_{11}}$};
% Text Node
\draw (155.41,35.83) node [anchor=north west][inner sep=0.75pt]    {$\Tilde{y}^{\mathrm{LP}}_{j} =1$};
% Text Node
\draw (194.38,245.83) node [anchor=north west][inner sep=0.75pt]    {$\mathbf{S_{21}}$};
% Text Node
\draw (194.38,289.28) node [anchor=north west][inner sep=0.75pt]    {$\mathbf{S_{31}}$};
% Text Node
\draw (334.66,35.83) node [anchor=north west][inner sep=0.75pt]    {$\Tilde{y}^{\mathrm{LP}}_{j} =0.5$};
% Text Node
\draw (190.45,375.45) node [anchor=north west][inner sep=0.75pt]    {$\mathbf{S_{K-1,1}}$};
% Text Node
\draw (192.24,418.9) node [anchor=north west][inner sep=0.75pt]    {$\mathbf{S_{K,1}}$};
% Text Node
\draw (278.79,376.17) node [anchor=north west][inner sep=0.75pt]    {$\mathbf{S_{K-1,2}}$};
% Text Node
\draw (286.34,151.69) node [anchor=north west][inner sep=0.75pt]    {$\mathbf{S_{12}}$};
% Text Node
\draw (329.79,151.69) node [anchor=north west][inner sep=0.75pt]    {$\mathbf{S_{13}}$};
% Text Node
\draw (286.34,245.83) node [anchor=north west][inner sep=0.75pt]    {$\mathbf{S_{22}}$};
% Text Node
\draw (286.34,289.28) node [anchor=north west][inner sep=0.75pt]    {$\mathbf{S_{32}}$};
% Text Node
\draw (329.79,245.83) node [anchor=north west][inner sep=0.75pt]    {$\mathbf{S_{23}}$};
% Text Node
\draw (329.79,289.28) node [anchor=north west][inner sep=0.75pt]    {$\mathbf{S_{33}}$};
% Text Node
\draw (407.14,150.97) node [anchor=north west][inner sep=0.75pt]    {$\mathbf{S_{1,K-1}}$};
% Text Node
\draw (458.72,150.97) node [anchor=north west][inner sep=0.75pt]    {$\mathbf{S_{1,K}}$};
% Text Node
\draw (408.14,246.55) node [anchor=north west][inner sep=0.75pt]    {$\mathbf{S_{2,K-1}}$};
% Text Node
\draw (270.48,64.79) node [anchor=north west][inner sep=0.75pt]    {$j_{1}$};
% Text Node
\draw (313.93,64.79) node [anchor=north west][inner sep=0.75pt]    {$j_{2}$};
% Text Node
\draw (394.41,64.07) node [anchor=north west][inner sep=0.75pt]    {$j_{K-2}$};
% Text Node
\draw (437.14,64.07) node [anchor=north west][inner sep=0.75pt]    {$j_{K-1}$};
% Text Node
\draw (46,126.34) node [anchor=north west][inner sep=0.75pt]    {$\Tilde{x}^{\mathrm{LP}}_{i} =1$};
% Text Node
\draw (34,311) node [anchor=north west][inner sep=0.75pt]    {$\Tilde{x}^{\mathrm{LP}}_{i} =0.5$};
% Text Node
\draw (107.03,223.38) node [anchor=north west][inner sep=0.75pt]    {$i_{1}$};
% Text Node
\draw (107.03,266.83) node [anchor=north west][inner sep=0.75pt]    {$i_{2}$};
% Text Node
\draw (98.62,353.72) node [anchor=north west][inner sep=0.75pt]    {$i_{K-2}$};
% Text Node
\draw (98,396) node [anchor=north west][inner sep=0.75pt]    {$i_{K-1}$};
% Text Node
\draw (107.76,439.17) node [anchor=north west][inner sep=0.75pt]    {$i_{K}$};
% Text Node
\draw (485.28,64.07) node [anchor=north west][inner sep=0.75pt]    {$j_{K}$};

\end{tikzpicture}

\caption{Illustration of the bundle partition}.
\label{fig:Sij}
\end{figure}

We first construct a partition of bundles. Denote 
\begin{align*}
 & \mathbf{N_1}=\{i\in \mathbf{N} \mid \Tilde{x}^{\mathrm{LP}}_i=1\}\cup\{0\},\ \mathbf{N_2}=\{i\in \mathbf{N} \mid \Tilde{x}^{\mathrm{LP}}_i=0.5\},\\ 
 & \mathbf{M_1}=\{j \in  \mathbf{M}  \mid \Tilde{y}^{\mathrm{LP}}_j=1\}\cup\{0\},\  \mathbf{M_2}=\{j\in \mathbf{M} \mid \Tilde{y}^{\mathrm{LP}}_j=0.5\} .  
\end{align*}
Let $ b_1\geq b_2\geq \cdots\geq b_{K-1} \geq b_K=0$ be a sequence of thresholds and $i_k=\max\{i\in \mathbf{N}\mid p_i\geq b_kr^*\},\ j_k=\max\{j\in \mathbf{M}\mid q_j\geq b_kr^*\},\ k=1,...,K$ be cutoff indexes, where $r^*$ is the optimal objective value of problem \eqref{prob:LP}. We partition bundles $(\mathbf{N_1}\cup \mathbf{N_2})\times (\mathbf{M_1}\cup \mathbf{M_2})$ with blocks defined as follows.
\begin{align*}
    \mathbf{S_{00}}&= \mathbf{N_1}\times \mathbf{M_1},\\
    \mathbf{S_{11}}&=\{(i,j)\in \mathbf{N_1}\times \mathbf{M_2} \mid j\leq j_1\}\bigcup\{(i,j)\in \mathbf{N_2}\times \mathbf{M_1} \mid i\leq i_1\}\bigcup \{(i,j)\in \mathbf{N_2}\times \mathbf{M_2} \mid i\leq i_1,\ j\leq j_1\},\\
    \mathbf{S_{1J}}&= \{(i,j)\in \mathbf{N_1}\times \mathbf{M_2} \mid j_{J-1}<j\leq j_{J}\}\bigcup \{(i,j)\in \mathbf{N_2}\times \mathbf{M_2} \mid i\leq i_1,\ j_{J-1}<j\leq j_{J}\},\ \forall J=2,\dots ,K,\\
    \mathbf{S_{I1}}&=\{(i,j)\in \mathbf{N_2}\times \mathbf{M_1} \mid i_{I-1}<i\leq i_{I}\}\bigcup \{(i,j)\in \mathbf{N_2}\times \mathbf{M_2}\mid i_{I-1}<i\leq i_{I},\ j\leq j_1\},\ \forall I=2,\dots, K,\\
    \mathbf{S_{IJ}}&=\{(i,j)\in \mathbf{N_2}\times \mathbf{M_2} \mid i_{I-1}<i\leq i_I,\ j_{J-1}<j\leq j_J\},\ \forall I,J=2,\dots, K. 
\end{align*}
Figure \ref{fig:Sij} provides an illustration of the above blocks. In this figure, rows (columns) represent products in category 1 (category 2) including the no-purchasing option. For illustration purpose, we remove all rows and columns with $\Tilde{x}^{\mathrm{LP}},\Tilde{y}^{\mathrm{LP}}=0$, and put the bundles in $\mathbf{S_{00}}$ at the top left corner. 

We consider $K$ candidate integral solutions (or assortments) 
\begin{align*}
    & \mathbf{S_k}=\mathbf{S_{00}}\bigcup\left(\bigcup_{I=1}^{k}\bigcup_{J=1}^{K+1-k}\mathbf{S_{IJ}}\right),\ k=1,...,K.
\end{align*}
As an illustration, in Figure \ref{fig:Sij}, the bundles in the red and blue rectangles represents $\mathbf{S_2}$ and $\mathbf{S_K}$, respectively. 
The following result shows that bundles with nonzero variables in the LP optimal solution are covered by the $K$ assortments under some conditions of the thresholds.
\begin{lemma}\label{lemma:cover}
If $b_k+b_{K-k}\leq 1$ for all $k=1,\dots,K-1$, then
$$\{(i,j)\mid \Tilde{z}^{\mathrm{LP}}_{ij}\neq 0\}\cup\{(i,0)\mid \Tilde{x}^{\mathrm{LP}}_i\neq 0\}\cup\{(0,j)\mid \Tilde{y}^{\mathrm{LP}}_j\neq 0\}\subseteq \cup_{k=1}^K \mathbf{S_k}.$$
\end{lemma}
\proof{Proof of Lemma \ref{lemma:cover}.}
This lemma follows directly from Lemma \ref{lemma:half-zij}. First, we observe all the decision variables $\Tilde{x}^{\mathrm{LP}},\Tilde{y}^{\mathrm{LP}},\Tilde{z}^{\mathrm{LP}}=1$ are covered by $S_{00}$. Then, by Lemma \ref{lemma:half-zij}, all the half-integral decision variables are covered by some $\mathbf{S_k}$.
\Halmos
\endproof

\subsubsection*{Step II: lower bound of approximation ratio.}
In this step, we derive a lower bound of the approximation ratio of the $K$ candidate assortments defined in the last step.
For each block $\mathbf{S_{IJ}}$, define 
\begin{align*}
    U_{IJ}&=\sum_{(i,j)\in\mathbf{S_{IJ}}}u_{ij},\ R_{IJ}=\sum_{(i,j)\in\mathbf{S_{IJ}}}(p_i+q_j)u_{ij}/U_{IJ},
\end{align*}
as the total preference weight and weighted average revenue, respectively. The revenue of each assortment $\mathbf{S_{k}},\ k=1,\dots,K$ can then be written as
\begin{equation*}
    R_k=\frac{U_{00}R_{00}+\sum_{I=1}^{k}\sum_{J=1}^{K+1-I}U_{IJ}R_{IJ}}{U_{00}+\sum_{I=1}^{k}\sum_{J=1}^{K+1-I}U_{IJ}}.
\end{equation*}

For $I,J=1,...,K$, define $$\mathbf{S'_{IJ}}=\mathbf{S_{IJ}}\setminus \{ (i,j) \mid z ^{\LP}_{ij}=0\}$$
as bundles in $\mathbf{S_{IJ}}$ with nonzero variables in the LP optimal solution and $U'_{IJ}$ and $R'_{IJ}$ as its total preference weight and weighted average revenue, respectively.

\begin{lemma}\label{lemma:Sij bounds}
For any $I,J=2,\dots, K$, it holds that
\begin{align}
\begin{split}\label{ineq:bounds Rij}
    R'_{11}&\geq b_1r^*,\\
    R'_{1J}&\geq b_Jr^*,\\
    R'_{I1}&\geq b_Ir^*,\\
    (b_{I-1}+b_{J-1})r^*&\geq R'_{IJ}\geq (b_I+b_J)r^*.
    \end{split}
\end{align}
\end{lemma}
\proof{Proof of Lemma \ref{lemma:Sij bounds}.} The proof  follows directly from the construction of $\mathbf{S'_{IJ}}$.
% Basically, we have that for each $(I,J)$ where $I=1$ or $J=1$, at least one of $p_i,q_j$ is lower bounded, by the definition of the cutoff point $i_1,\dots,i_K,j_1,\dots,j_K$. Note that the lower bound on the value of $p_i,q_j$'s where $x_i=1/2,y_j=1/2$. are sufficient to generate the bound. For the upper and lower bound of $R_{IJ}$ where $I\neq 1$ and $J\neq 1$, note that all items or bundles $(i,j)\in\mathbf{S_{IJ}}$ are those with $z_{ij}=1/2$, and their $p_i,q_j$ are bounded by $b_Ir^*,b_{I-1}r^*,b_{J}^*,b_{J-1}r^*$, so it directly follows from the construction of cutoff point $i_1,\dots,i_K,j_1,\dots,j_K$.
\Halmos
\endproof
Lemma \ref{lemma:Sij bounds} can be extended to the weighted average revenue of any subset of $\mathbf{S'_{IJ}}$.
The following result provides a lower bound of the approximation ratio of each candidate assortment.
\begin{lemma}\label{lemma:ratio_lower_bound}
For each $k=1,...,K$, we have
\begin{align*}
R_k\geq \min\left\{\frac{U_{00}R_{00} +\sum_{I=1}^{k}\sum_{J=1}^{K+1-I}U'_{IJ}R'_{IJ}}{U_{00}+\sum_{I=1}^{k}\sum_{J=1}^{K+1-I}U'_{IJ}},(b_k+b_{K+1-k})r^*\right\}.
\end{align*}
\end{lemma}
\proof{Proof of Lemma \ref{lemma:ratio_lower_bound}.}
Denote $\mathbf{S''_{IJ}}=\mathbf{S_{IJ}}\setminus\mathbf{S'_{IJ}}$, and define $R''_{IJ}$ and $U''_{IJ}$ accordingly. It follows that
\begin{align*}
    R_k&=\frac{U_{00}R_{00}+\sum_{I=1}^{k}\sum_{J=1}^{K+1-I}U_{IJ}R_{IJ}}{U_{00}+\sum_{I=1}^{k}\sum_{J=1}^{K+1-I}U_{IJ}}\notag\\
    &=\frac{U_{00}R_{00}+\sum_{I=1}^{k}\sum_{J=1}^{K+1-I}U'_{IJ}R'_{IJ}+\sum_{I=1}^{k}\sum_{J=1}^{K+1-I}U''_{IJ}R''_{IJ}}{U_{00}+\sum_{I=1}^{k}\sum_{J=1}^{K+1-I}U'_{IJ}+\sum_{I=1}^{k}\sum_{J=1}^{K+1-I}U''_{IJ}}\notag\\
    &\geq\min\left\{\frac{U_{00}R_{00}+\sum_{I=1}^{k}\sum_{J=1}^{K+1-I}U'_{IJ}R'_{IJ}}{U_{00}+\sum_{I=1}^{k}\sum_{J=1}^{K+1-I}U'_{IJ}},\ \frac{\sum_{I=1}^{k}\sum_{J=1}^{K+1-I}U''_{IJ}R''_{IJ}}{\sum_{I=1}^{k}\sum_{J=1}^{K+1-I}U''_{IJ}}\right\}\\
    &\geq\min
    \left\{\frac{U_{00}R_{00}+\sum_{I=1}^{k}\sum_{J=1}^{K+1-I}U'_{IJ}R'_{IJ}}{U_{00}+\sum_{I=1}^{k}\sum_{J=1}^{K+1-I}U'_{IJ}},\ (b_k+b_{K+1-k})r^*\right\}.
\end{align*}
\Halmos
\endproof

\subsubsection*{Step III: approximation ratio optimization.}
Lemma \ref{lemma:ratio_lower_bound} implies that the $K$ candidate assortments guarantee an approximation ratio of $\min\{\beta,\max_{k=1,...,K}\{(b_k+b_{K+1-k})\}\}$, where $\beta$ satisfies
\begin{align*}
  \max_{k=1,\dots,K}\left\{\frac{U_{00}R_{00}+\sum_{I=1}^{k}\sum_{J=1}^{K+1-I}U'_{IJ}R'_{IJ}}{U_{00}+\sum_{I=1}^{k}\sum_{J=1}^{K+1-I}U'_{IJ}}\right\}\geq \beta r^*.   
\end{align*}
If the assumption in Lemma \ref{lemma:cover} holds, then we can write the optimal objective value of the LP as
\begin{equation*}
    r^*=\frac{U_{00}R_{00}+0.5\sum_{I=1}^{K}\sum_{J=1}^{K+1-I}U'_{IJ}R'_{IJ}}{U_{00}+0.5\sum_{I=1}^{K}\sum_{J=1}^{K+1-I}U'_{IJ}},
\end{equation*}
which allows for the following quadratic program to compute the largest $\beta$ in the worst case of $U_{00},R_{00},U_{IJ}',R_{IJ}'$.
\begin{subequations}
\label{eq:find beta}
\begin{align}
    \beta(b)=&\min_{U_{00},R_{00},U',R',\beta} \beta \notag\\
    \text{s.t. } &U_{00}R_{00}+0.5\sum_{I=1}^{K}\sum_{J=1}^{K+1-I}U'_{IJ}R'_{IJ}=U_{00}+0.5\sum_{I=1}^{K}\sum_{J=1}^{K+1-I}U'_{IJ}  \label{eq:find_beta_1}\\ 
    &U_{00}R_{00}+\sum_{I=1}^{k}\sum_{J=1}^{K+1-k}U'_{IJ}R'_{IJ}\leq \beta\left(U_{00}+\sum_{I=1}^{k}\sum_{J=1}^{K+1-k}U'_{IJ}\right),\ k=1,\dots,K     \label{eq:find_beta_2}\\ 
    & R_{00}\geq 0,\ R'_{11}\geq b_1,\ R'_{1J}\geq b_J,\ R'_{I1}\geq b_I,\ (b_{I-1}+b_{J-1})\geq R'_{IJ}\geq (b_I+b_J),\forall I,J=2,...,K \label{eq:find_beta_4}\\ 
    & U_{00}\geq 1,\ U'_{IJ} \geq 0,\ \forall I,J=1,...,K. \label{eq:find_beta_3}
\end{align}
\end{subequations}
Note that in problem \eqref{eq:find beta}, we normalize $R_{00}$ and all the $R'_{IJ}$ by $r^*$, and $U_{00}\geq 1$ is from $u_{00}=1$. Finally, we optimize the thresholds $b_1,...,b_{K-1}$ to obtain the tightest approximation ratio, which is referred to as $\alpha^*$, i.e.,
\begin{align}
    \begin{split}\label{eq:find alpha}
      \alpha^*= \max_b\ & \min\{\beta(b),\max_{k=1,...,K}\{(b_k+b_{K+1-k})\}\}\\
      \text{s.t. } & b_k+b_{K-k}\leq 1,\ \forall k=1,\dots,K-1\\
      & b_k\geq b_{k+1}\geq 0,\ \forall k=1,\dots,K-1.
    \end{split}
\end{align}

Summarizing the above three steps, we have the following result.
\begin{theorem}\label{thm:alpha*}
For a given positive integer $K$, we have
\begin{equation*}
    \max_{k=1,...,K}R_k \geq \alpha^* r^*
\end{equation*}
when $R_k,\ k=1,...,K$ is constructed with the optimal $b_1,...,b_{K-1}$ in problem \eqref{eq:find alpha}.
\end{theorem}

Theorem \ref{thm:alpha*} shows that the approximation framework yields an $\alpha^*$-approximation solution and this approximation ratio is clearly increasing in $K$. However, the optimization problems \eqref{eq:find beta} and \eqref{eq:find alpha} are intractable when $K$ is large. Interestingly, we show in the next subsection that we can achieve an approximation ratio close to the integrality gap even with a small value of $K=4,6$.

\subsection{Cases of $K=4,6$}\label{subsec:K46}
In this subsection, we provide a method to derive a lower bound of $\alpha^*$. Note that $\beta(b)$ is the supremum of $\beta$ such that constraints (\ref{eq:find_beta_1}-\ref{eq:find_beta_3}) are \textbf{infeasible}. Let $v=(v_1,...,v_K)\in \RR^K_+$ be a nonzero nonnagative vector. For any $\beta\in (0,1)$ and $U_{00},R_{00},U'_{IJ},R'_{IJ}$ satisfying bounds (\ref{eq:find_beta_4}-\ref{eq:find_beta_2}), define a function
\begin{align*}
    & f(\beta,U_{00},R_{00},U'_{IJ},R'_{IJ})\\
    \overset{\underset{\mathrm{def}}{}}{=} & 2U_{00}+\sum_{I=1}^{K}\sum_{J=1}^{K+1-I}U'_{IJ}-\left(2U_{00}R_{00}+\sum_{I=1}^{K}\sum_{J=1}^{K+1-I}U'_{IJ}R'_{IJ}\right)\\
 - & \sum_{k=1}^K v_k\left[\beta\left(U_{00}+\sum_{I=1}^{k}\sum_{J=1}^{K+1-k}U'_{IJ}\right)-U_{00}R_{00}-\sum_{I=1}^{k}\sum_{J=1}^{K+1-k}U'_{IJ}R'_{IJ} \right]\\
 =& U_{00}\left(2(1-R_{00})+(R_{00}-\beta)\sum_{k=1}^K v_k\right)+\sum_{I=1}^K\sum_{J=1}^{K+1-I}U'_{IJ}\left((1-R'_{IJ})+(R'_{IJ}-\beta)\sum_{k=I}^{K+1-J}v_k\right).
\end{align*}
It is clear that $f(\beta,U_{00},R_{00},U'_{IJ},R'_{IJ})$ is strictly decreasing in $\beta$ as $U_{00}\geq 1$ and $v$ is nonzero. 
If $f(\beta,U_{00},R_{00},U'_{IJ},R'_{IJ})>0$ for any  $U_{00},R_{00},U'_{IJ},R'_{IJ}$ satisfying bounds (\ref{eq:find_beta_4}-\ref{eq:find_beta_2}), then at least one constraint in (\ref{eq:find_beta_1}-\ref{eq:find_beta_2}) is violated, which implies that constraints (\ref{eq:find_beta_1}-\ref{eq:find_beta_3}) are infeasible for this $\beta$ and thus $\beta<\beta(b)$. 
Therefore, a lower bound $\beta'$ of $\beta(b)$ can be obtained by constructing a dual vector $v$ such that $f(\beta',U_{00},R_{00},U'_{IJ},R'_{IJ})\geq 0$ for any  $U_{00},R_{00},U'_{IJ},R'_{IJ}$ satisfying bounds (\ref{eq:find_beta_4}-\ref{eq:find_beta_2}). 
This condition is satisfied if and only if all the coefficients of $U_{00},U_{IJ}'$ in $f(\beta',U_{00},R_{00},U'_{IJ},R'_{IJ})$ are nonnegative for all $R_{00},R_{IJ}'$ satisfying  \eqref{eq:find_beta_4}. Since there is no upper bound for $R_{00},R'_{1J},R'_{I1}$ in \eqref{eq:find_beta_4}, their coefficients should be nonnegative, i.e.,
\begin{align}\label{eq:coef1}
    \sum_{k=1}^Kv_k-2\geq 0,\     \sum_{k=1}^{K+1-J}v_k -1\geq 0,\ \sum_{k=I}^{K} v_k-1\geq 0.
\end{align}
Moreover, given \eqref{eq:coef1}, the nonnegativity of the coefficients of $U_{00},U'_{IJ}$  is equivalent to 
\begin{subequations}\label{eq:coef2}
\begin{align}
& \min_{R_{00}\geq 0} 2(1-R_{00})+(R_{00}-\beta)\sum_{k=1}^K v_k =2-\beta'\sum_{k=1}^K v_k\geq 0,\label{eq:coef2_1}\\
 & \min_{R_{1J}'\geq b_J}  (1-R'_{1J})+(R'_{1J}-\beta)\sum_{k=1}^{K+1-J}v_k =1-b_J+(b_J-\beta') \sum_{k=1}^{K+1-J}v_k\geq 0,\ \forall J=1,...,K,\label{eq:coef2_2}\\
 & \min_{R_{I1}'\geq b_I}  (1-R'_{I1})+(R'_{I1}-\beta)\sum_{k=I}^{K}v_k =1-b_I+(b_I-\beta')\sum_{k=I}^{K} v_k\geq 0,\ \forall I=1,...,K,\label{eq:coef2_3}\\
 & \min_{R_{IJ}'}  (1-R'_{IJ})+(R'_{IJ}-\beta)\sum_{k=I}^{K+1-J}v_k, \quad \text{ s.t. } (b_{I-1}+b_{J-1})\geq R'_{IJ}\geq (b_I+b_J) \notag\\
 =& 1-\beta'\sum_{k=I}^{K+1-J}v_k+(b_{I-1}+b_{J-1})(\sum_{k=I}^{K+1-J}v_k-1)-(b_{I-1}+b_{J-1}-b_I-b_J)(\sum_{k=I}^{K+1-J}v_k-1)^+\geq 0,\notag\\
 & \forall I,J=2,...,K. \label{eq:coef2_4}
\end{align}
\end{subequations}

Therefore, if we can construct $\beta',b_1,...,b_{K-1},v_1,...,v_K$ such that \eqref{eq:coef1}\eqref{eq:coef2} and constraints of \eqref{eq:find alpha} are satisfied, then we get a lower bound of $\alpha^*$, i.e., $\min\{\beta',\max_{k=1,...,K}\{(b_k+b_{K+1-k})\}\}$.

\begin{theorem}\label{thm:apx_ratio}
\begin{enumerate}[label=(\alph*)] 		%\arabic \roman \Roman \Alpha
  \item If $K=4$, then $\alpha^*\geq \frac{5+\sqrt{5}}{10}\approx 0.7236$.
  \item If $K=6$, then $\alpha^*\geq 0.74$.
\end{enumerate}
\end{theorem}

\proof{Proof of Theorem \ref{thm:apx_ratio}.}
(a)
Let $\beta'=b_1=\frac{5+\sqrt{5}}{10},\ b_2 = \frac{\sqrt{5}}{5},\ b_3=\frac{5-\sqrt{5}}{10},\ v_1=v_4=1,\ v_2=v_3=\frac{b_1+b_2-1}{b_1+b_2-\beta'}=\frac{3-\sqrt{5}}{2}\approx 0.38$. Since $v_1=v_4=1$, \eqref{eq:coef1} clearly holds. By calculation, one can check the constraints of \eqref{eq:find alpha} and \eqref{eq:coef2} are satisfied. Hence, $\alpha^*\geq \min\{\beta',b_1,b_2+b_3\}=\frac{5+\sqrt{5}}{10}$.

(b) Let $\beta'=0.74$,  $b_1=0.74,b_2=0.484,b_3=0.399,b_4=0.341,b_5=0.256$,
$v_1=v_6=1, v_2=v_5\approx 0.23772, v_3=v_4\approx 0.11264$. It can be checked that \eqref{eq:coef1}\eqref{eq:coef2} and constraints of \eqref{eq:find alpha} are satisfied. Hence, $\alpha^*\geq \min\{\beta',b_1,b_2+b_5,b_3+b_4\}=0.74$.
% The details of the computation process are redirected to Section \ref{section:K6lower} in the electric companion.
\Halmos
\endproof

\begin{corollary}
It holds that $0.74\leq \Gap\leq 0.75$.
\end{corollary}

It is worthy mentioning how the values of  $\beta',b_1,...,b_{K-1},v_1,...,v_K$ are constructed in the proof of Theorem \ref{thm:apx_ratio}.
To satisfy \eqref{eq:coef1}, we set $v_1=v_K=1$. Moreover, we set $v_k=v_{K+1-k},\ k=1,...,K$. To ensure $\beta'$ is large, we set $2-\beta'\sum_{k=1}^K v_k=0$, otherwise $f(\beta',\cdots)>0$ as $U_{00}\geq 1$ and we can increase $\beta'$. 
% When $K=4$, plugging in the bounds on $R_{00},R'_{IJ}$ implied by \eqref{eq:find_beta_4}, as well as the constraints of \eqref{eq:find alpha}, one can verify that inequalities \eqref{eq:coef2} reduce to the following constraints.  NO NEED TO USE 7c and 8
When $K=4$, only the values of $v_2=v_3$ and $\beta'$ are to be determined. We further set the inequality of \eqref{eq:coef2_4} for $I=2,J=3$ to be equality, i.e.,
\begin{align*}
    1-\beta'v_2+(b_1+b_2)(v_2-1)-(b_1-b_3)(v_2-1)^+=0.
\end{align*}
We have $v_2<1$ as otherwise $1-\beta'v_2=-(b_2+b_3)(v_2-1)<0$, contradicting $2-\beta'(2+v_2+v_3)=0$.
Thus, we have the following equations:
%fix the value of $b_1,b_2,b_3$ in a grid of the feasible set $\{(b_1,b_2,b_3)\mid b_1+b_3\leq 1,\ b_1\geq b_2\geq b_3\geq 0\}$, and then set one of the inequalities in \eqref{eq:coef2} to be equality (besides the first one). By a grid search of $b$ and brute forcing all the possible equations, we find that when $b_1 \approx \frac{5+\sqrt{5}}{10},\ b_2 \approx \frac{\sqrt{5}}{5},\ b_3\approx\frac{5-\sqrt{5}}{10}$ and the coefficient of $U'_{23}$ (or $U'_{32}$) is tight, the solution of
\begin{align*}
   v_1=v_4=1,\ v_2=v_3,\ 2-\beta'\sum_{k=1}^4v_k=0,\  (1-b_1-b_2)+(b_1+b_2-\beta')v_2=0.
\end{align*}
Solving the above equations yields 
$$v_1=v_4=1,v_2=v_3=\frac{b_1+b_2-1}{b_1+b_2-\beta'},\beta'=b_1+b_2-\sqrt{(b_1+b_2)^2-b_1-b_2},\ \text{ if } b_1+b_2\geq 1.$$
One can check that all the inequalities of \eqref{eq:coef2} are satisfied with this solution if $b_1+b_2\geq 1$.
% Note that the above equations implies a quadratic equation $\beta'(1-b_1-b_2)+(b_1+b_2-\beta')(1-\beta')=0$ and thus a closed form solution of $\beta'(b)=b_1+b_2-\sqrt{(b_1+b_2)^2-b_1-b_2}$ when $b_1+b_2\geq 1$. 
The values of $b_1,b_2,b_3$ in the proof of Theorem \ref{thm:apx_ratio} can be calculated by plugging the above solution into problem \eqref{eq:find alpha} and solving the maximization problem. The optimal solution for \eqref{eq:find alpha} is $\beta'=b_1 = \frac{5+\sqrt{5}}{10},\ b_2 = \frac{\sqrt{5}}{5}$, and $b_3=\beta'-b_2=\frac{5-\sqrt{5}}{10}$. 
% However, we cannot prove that $\beta(b)=b_1+b_2-\sqrt{(b_1+b_2)^2-b_1-b_2}$ is the optimal objective value in \eqref{eq:find beta}. 
For the case of $K=6$, a similar method would imply that $\beta'$ is the solution of a cubic equation while we cannot obtain a closed-form solution to the value of $b_1,...,b_6$ for the maximization problem \eqref{eq:find alpha}. Instead, we conduct a grid search over the feasible region of $b_1,...,b_6$ implied by the constraints in \eqref{eq:find alpha} to obtain a set of threshold values that gives the approximation factor of 0.74.

Theorem \ref{thm:apx_ratio} shows that one of the six candidate assortments $S_k,k=1,...,6$ with the thresholds given in the proof has an approximation ratio within 0.01 of the integrality gap of the LP relaxation. When we increase $K$, the approximation ratio would be closer to the integrality gap. However, problems \eqref{eq:find beta} and \eqref{eq:find alpha} are not easy to deal with when $K$ is large. In the next subsection, we modify the approximation framework to obtain a $(\Gap-\varepsilon)$-approximation algorithm but with a time complexity depending exponentially on $\varepsilon$.

\subsection{A ($\Gap-\varepsilon$)-approximation scheme}\label{subsec:gap-e}

In this subsection, we modify the approximation framework to achieve an approximation ratio within $\varepsilon$ of the integrality gap of the LP relaxation for any $\varepsilon>0$. 

The modified algorithm consists of two steps. First, solve problem \eqref{prob:LP} and obtain an optimal solution and the optimal objective $r^*$. If the optimal solution is non-integral, we use specific thresholds $b_k=1-k\varepsilon,k=0,1,...,\frac{1}{\varepsilon}$ to construct blocks of bundles for products $i,j$ with $x_i ^{\LP},y ^{\LP}_j\neq 0$. A block is a set of products or bundles, whose prices $p_i,q_j$ are upper and lower bounded by some thresholds in $b_k$'s. Second, we compare the revenue of all assortments of the blocks (i.e. all $x_i,y_j,z_{ij}$ are set to be the same within one block) and output the largest one. We prove that the output assortment guarantees an approximation ratio within $\varepsilon$ of the integrality gap. 
% In other words, we consider another problem instance that regards a block as one product or bundle. 
The second step can be regarded as an assortment optimization for an instance with $\frac{1}{\varepsilon}$ products in both categories.
Different from considering only $K=\frac{1}{\varepsilon}$ special assortment of blocks in the partition-and-optimize approximation framework, we consider solutions corresponding to all possible combinations of constructed blocks. 

Without loss of generality, assume  $K=\frac{1}{\varepsilon}$ is a positive integer. Let $(\Tilde{x} ^{\LP},\Tilde{y} ^{\LP},\Tilde{z} ^{\LP})$ be a non-integral optimal solution of problem \eqref{prob:LP} divided by $w ^{\LP}$. Let $b_k=1-k\varepsilon,k=0,1,...,K$ be a sequence of thresholds.
The products are partitioned by
\begin{align*}
  \mathbf{N_{1,I}} &=\{i\in \mathbf{N_1}\setminus \{0\}\mid p_i\in [b_Ir^*,b_{I-1}r^*)\},\ I=1,...,K,  \\
  \mathbf{N_{2,I}}& =\{i\in \mathbf{N_2}\mid p_i\in [b_Ir^*,b_{I-1}r^*)\},\ I=1,...,K,\\
  \mathbf{M_{1,J}}& =\{j\in \mathbf{M_1}\setminus \{0\} \mid q_j\in [b_Jr^*,b_{J-1}r^*)\},\ J=1,...,K,\\
  \mathbf{M_{2,J}}& =\{j\in \mathbf{M_2}\mid q_j\in [b_Jr^*,b_{J-1}r^*)\},\ J=1,...,K,\\
  \mathbf{N_{\geq}} & = \{i\in N\mid p_i\geq r^*\},\ \mathbf{M_{\geq}} = \{j\in M\mid q_i\geq r^*\}.
\end{align*}
The ($\Gap-\varepsilon$)-approximation scheme is presented in Algorithm \ref{alg:gap-eps}.

\begin{algorithm}
\SetKwInOut{Input}{Input}
\SetKwInOut{Output}{Output}
\Input{$u_{ij},p_i,q_j$ for $i\in  \mathbf{N}_+ ,j\in  \mathbf{M}_+ $, $K=\frac{1}{\varepsilon}$}
Solve problem \eqref{prob:LP} to obtain $r^*$\\
Compute $\pi(x,y)$ of all integral solutions $(x,y)$ satisfying:
\begin{enumerate}[label=(\alph*)] 		%\arabic \roman \Roman \Alpha
  \item $x_i=1, \forall i\in \mathbf{N_{\geq}}$, $y_j=1,\forall j\in \mathbf{M_{\geq}}$,
    \item $x_i=x_k$ if $i,k\in \mathbf{N_{1,I}}$ or $i,k\in \mathbf{N_{2,I}}$ for some $I$,
    \item $y_j=y_l$ if $j,l\in \mathbf{M_{1,J}}$ or $j,l\in \mathbf{M_{2,J}}$  for some $J$,
    \item $x_i=0$ if $x ^{\LP}_i=0$ and $y_j=0$ if $y ^{\LP}_j=0$.
\end{enumerate}
\Output{the integral solution with the largest revenue}
\caption{}\label{alg:gap-eps}
\end{algorithm}

We next prove that Algorithm \ref{alg:gap-eps} is  a $(\Gap-4\varepsilon)$-approximation algorithm. The idea of the proof is to construct a problem instance $\mathcal{I}^{\varepsilon}$, from the original instance $\mathcal{I}$, such that 
\begin{itemize}
    \item the optimal objective values of the LP for both instances are close, i.e., $r^*_{\mathcal{I}}\approx r^*_{\mathcal{I}^{\varepsilon}}$ and
    \item the optimal objective values of the assortment optimization problem \eqref{prob:IP} for both instances are close, i.e., $\pi^*_{\mathcal{I}}\approx\pi^*_{\mathcal{I}^{\varepsilon}}$.
    % \item The optimal assortment of instance $\mathcal{I}^{\varepsilon}$ can be found efficiently for any size $m,n$ of instance $\mathcal{I}$ and fixed $\varepsilon$. NO NEED TO MENTION EFFICIENCY AS THE COMPLEXITY IS EXPONENTIAL IN EPSILON
\end{itemize}

\begin{theorem}\label{thm:gap-eps}
For any $\varepsilon>0$, Algorithm \ref{alg:gap-eps} is a $(\Gap-4\varepsilon)$-approximation algorithm.
\end{theorem}
\proof{Proof of Theorem \ref{thm:gap-eps}.}
We only prove the case  $\mathbf{N_{\geq}}=\mathbf{M_{\geq}}=\emptyset$ for simplicity, as the general case follows the same argument. The problem instance is denoted by $\mathcal{I}$.
Define blocks as
\begin{align*}
    \mathbf{S^{st}_{IJ}}&= \mathbf{N_{s,I}}\times \mathbf{M_{t,J}},\ s,t\in\{1,2\} ,\ I,J\in\{1,\dots,K\},\\
    \mathbf{S^{0t}_{J}}&=\{0\}\times \mathbf{M_{t,J}},\ J\in \{1,...,K\}\\
    \mathbf{S^{s0}_{I}}&=\mathbf{N_{s,I}}\times \{0\},\ I\in\{1,...,K\}.
\end{align*} 
Define $U^{st}_{IJ},U^{0t}_J,U^{s0}_I$ as the total preference weight and $R^{st}_{IJ},R^{0t}_J,R^{s0}_I$ as the weighted average revenue for block $\mathbf{S_{IJ}^{st}},\mathbf{S^{0t}_J},\mathbf{S^{s0}_I}$, respectively.
We create a new problem instance $\mathcal{I}^{\varepsilon}$ as follows. Category 1 has $2K$ products labeled by $i_{s,I}$, $s=1,2$, $I=1,...,K$, and the no-purchasing option is labeled by $i_0$. Category 2 has $2K$ products labeled by $j_{t,J}$, $t=1,2$, $J=1,...,K$, and the no-purchasing option is labeled by $j_0$. The price of product $i_{s,I}$ ($j_{t,J}$) is any price of product in $N_{s,I}$ ($M_{t,J})$ and is denoted by $p_{s,I}$ ($q_{t,J}$). The prices of $i_0,j_0$ are zero.
The preference weight of bundle $(i_{s,I},j_{t,J}),(i_{s,I},j_0),(i_0,j_{t,J})$ is  $U_{IJ}^{st},U^{s0}_I,U^{0t}_J$, respectively.
Clearly, we have
\begin{align}\label{eq:p+q bound}
\begin{aligned}
 |p_{s,I}+q_{t,J}-R_{IJ}^{st}| & \leq 2\varepsilon r^*,\\
 |p_{s,I}-R^{s0}_I| & \leq \varepsilon r^*,\\
 |q_{t,J}-R^{0t}_J| & \leq \varepsilon r^*,\ \forall s,t\in \{1,2\}, I,J\in\{1,...,K\}.
\end{aligned}
\end{align}

Let $(x^{\varepsilon},y^{\varepsilon})$ be an optimal solution of the instance $\mathcal{I}^{\varepsilon}$. We can construct a feasible solution $(x',y')$ of instance $\mathcal{I}$ by setting $x'_i=x^{\varepsilon}_{i_{s,I}}$ if $i\in \mathbf{N_{s,I}}$ and $y'_j=y^{\varepsilon}_{j_{t,J}}$ if $j\in \mathbf{M_{t,J}}$, otherwise $x'_i=y'_j=0$. Note that $(x',y')$ is one of the integral solutions evaluated in Algorithm \ref{alg:gap-eps}. It follows that the optimal revenue of instance $\mathcal{I}^{\varepsilon}$ satisfies
\begin{align*}
\pi_{\mathcal{I}^\varepsilon}(x^{\varepsilon},y^{\varepsilon})= & \frac{\sum  U_{IJ}^{st}(p_{i_{s,I}}+q_{j_{t,J}})x^{\varepsilon}_{i_{s,I}}y^{\varepsilon}_{j_{t,J}}+\sum U^{s0}_Ip_{i_{s,I}}x^{\varepsilon}_{i_{s,I}}+\sum U^{0t}_J q_{j_{t,J}}y^{\varepsilon}_{j_{t,J}}}{1+\sum  U_{IJ}^{st}x^{\varepsilon}_{i_{s,I}}y^{\varepsilon}_{j_{t,J}}+\sum U^{s0}_I x^{\varepsilon}_{i_{s,I}}+\sum U^{0t}_J y^{\varepsilon}_{j_{t,J}}}\\
\leq & \frac{\sum U_{IJ}^{st}R^{st}_{IJ}x^{\varepsilon}_{i_{s,I}}y^{\varepsilon}_{j_{t,J}}+\sum U^{s0}_I R^{s0}_I x^{\varepsilon}_{i_{s,I}}+\sum U^{0t}_J R^{0t}_J y^{\varepsilon}_{j_{t,J}}}{1+\sum  U_{IJ}^{st}x^{\varepsilon}_{i_{s,I}}y^{\varepsilon}_{j_{t,J}}+\sum U^{s0}_I x^{\varepsilon}_{i_{s,I}}+\sum U^{0t}_J y^{\varepsilon}_{j_{t,J}}}+2\varepsilon r^*\\
= & \pi_{\mathcal{I}}(x',y')+2\varepsilon r^*.    
\end{align*}

In the following, we show that the optimal objective value of the LP relaxation \eqref{prob:LP} of instance $\mathcal{I}^{\varepsilon}$ is no less than $(1-2\varepsilon)r^*$.
From the optimal solution $(\Tilde{x}^{\mathrm{LP}},\Tilde{y}^{\mathrm{LP}},\Tilde{z}^{\mathrm{LP}})$ of the LP relaxation of instance $\mathcal{I}$, we construct a feasible solution $(x^\varepsilon,y^\varepsilon,z^\varepsilon)$ of the LP relaxation of problem instance $\mathcal{I}^{\varepsilon}$ as follows.
We set $x_{i_{1,I}}=y_{j_{1,J}}=1$ and $x_{i_{2,I}}=y_{j_{2,J}}=0.5$. 
For $z_{i_{2,I}j_{2,J}}$ not uniquely determined by the constraints of the LP, we set it to be $0$ if all $z_{ij}^{\mathrm{LP}}$ in block $\mathbf{S^{22}_{IJ}}$ are equal to zero, otherwise, we set it to be $0.5$. 
If all the $z^{\mathrm{LP}}_{ij}\in\mathbf{S^{22}_{IJ}}$ are not equal, then we have $b_{I-1}+b_{J-1}\geq 1$ and $b_I+b_J\leq 1$ from Lemma \ref{lemma:half-zij}. By definition of the thresholds, we have $p_{i_{2,I}}+q_{j_{2,J}}\geq (b_I+b_J)r^*> (b_{I-1}+b_{J-1}-2\varepsilon)r^*\geq p_{i}+q_{j}-2\varepsilon r^*$ for any $(i,j)\in \mathbf{S^{22}_{IJ}}$. We denote the set of such blocks $\mathbf{S^{22}_{IJ}}$ where $z^{\mathrm{LP}}_{ij}$ for all $(i,j)\in \mathbf{S^{22}_{IJ}}$ are not equal by $\mathbf{S_{\Delta}}$.
It follows that the LP objective of the constructed solution $(x,y,z)$, denoted by $r^{\varepsilon}$, is no less than
\begin{align*}
& \frac{\sum_{\mathbf{S^{st}_{IJ}}\notin\mathbf{S_{\Delta}}} U_{IJ}^{st}R^{st}_{IJ} z_{i_{s,I}j_{t,J}}+\sum U^{s0}_I R^{s0}_I x^{\varepsilon}_{i_{s,I}}+\sum U^{0t}_J R^{0t}_J y^{\varepsilon}_{j_{t,J}}+0.5\sum  U_{IJ}'(b_{I-1}+b_{J-1})r^*}{1+\sum_{\mathbf{S^{st}_{IJ}}\notin\mathbf{S_{\Delta}}} U_{IJ}^{st}z_{i_{s,I}j_{t,J}}+\sum U^{s0}_I x^{\varepsilon}_{i_{s,I}}+\sum U^{0t}_J y^{\varepsilon}_{j_{t,J}}+0.5\sum U_{IJ}'}-2\varepsilon r^*\\
\geq & \frac{\sum_{\mathbf{S^{st}_{IJ}}\notin\mathbf{S_{\Delta}}} U_{IJ}^{st}R^{st}_{IJ} z_{i_{s,I}j_{t,J}}+\sum U^{s0}_I R^{s0}_I x^{\varepsilon}_{i_{s,I}}+\sum U^{0t}_J R^{0t}_J y^{\varepsilon}_{j_{t,J}}+\sum \sum_{\mathbf{S_{IJ}}\in \mathbf{S_{\Delta}}} u_{ij}(p_i+q_j)z_{ij} ^{\LP}}{1+\sum_{\mathbf{S^{st}_{IJ}}\notin\mathbf{S_{\Delta}}} U_{IJ}^{st}z_{i_{s,I}j_{t,J}}+\sum U^{s0}_I x^{\varepsilon}_{i_{s,I}}+\sum U^{0t}_J y^{\varepsilon}_{j_{t,J}}+\sum \sum_{\mathbf{S_{IJ}}\in \mathbf{S_{\Delta}}} u_{ij}z_{ij} ^{\LP}}-2\varepsilon r^*\\
=& r^*-2\varepsilon r^*.
\end{align*}

Summarizing the above results, we have 
$$\frac{\pi_{\mathcal{I}}(x',y')}{r^*}\geq \frac{\pi_{\mathcal{I^\varepsilon}}(x^{\varepsilon},y^{\varepsilon})-2\varepsilon r^*}{r^*}\geq \frac{\pi_{\mathcal{I}^\varepsilon}(x^{\varepsilon},y^{\varepsilon})}{r^{\varepsilon}}(1-2\varepsilon)-2\varepsilon\geq (1-2\varepsilon)\Gap-2\varepsilon\geq \Gap- 4\varepsilon.$$

\Halmos
\endproof

Algorithm \ref{alg:gap-eps} needs to find the largest revenue of $2^{4/\varepsilon}$ number of integral solutions, its time complexity is $O(\exp(\frac{1}{\varepsilon}))$. To outperform the 0.74-approximation algorithm in the previous subsection, it requires $\varepsilon=\frac{1}{40}$, which is time-consuming.
\begin{remark}\label{remark:general price}
Algorithm \ref{alg:gap-eps} can be extended to models with more general bundle price structures. Consider the pricing scheme with the price of the bundle $(i,j)$ to be $f(c_1p_i+c_2q_j)$ for some Lipschitz continuous function with Lipschitz constant $C$, positive constants $c_1,c_2$ and satisfies $f(c_1p_i+c_2q_j)\geq\max\{p_i,q_j\}$. 
% The inequality condition is to ensure that the 
Such pricing scheme captures many interesting cases in practice such as a percentage discount with $f(x)=Cx,c_1=c_2=1,C\in[\max_{i,j}\{\frac{p_i}{p_i+q_j},\frac{q_j}{p_i+q_j}\},1)$ and a constant discount with $f(x)=x-a,c_1=c_2=1,0<a\leq\min_{i,j}\{p_i,q_j\}$. The proof of the approximation ratio of Algorithm \ref{alg:gap-eps} under this pricing scheme is similar to that of Theorem \ref{thm:gap-eps}, with the first inequality in \eqref{eq:p+q bound} modified to 
    \[
    |f(c_1p_{s,I}+c_2q_{t,J})-R_{IJ}^{st}| \leq 2C\max\{c_1,c_2\}\varepsilon r^*,
    \]
    and the rest of the proof accordingly. The approximation ratio of Algorithm \ref{alg:gap-eps} would be $\Gap_f-4C\max\{c_1,c_2\}\varepsilon$, where $\Gap_f$ is the integrality gap of the LP relaxation under this pricing scheme. 
    
    % In some cases, the price of bundles $(i,j)$ might not be exactly $p_i+q_j$, but a function $f(p_i+q_j)$. For any $f(\cdot)$ that is Lipschitz continuous with Lipschitz constant $C$ and satisfy $f(p_i+q_j)\geq\max\{p_i,q_j\}$, Algorithm \ref{alg:gap-eps} can be directly applied and still have an approximation guarantee that can be arbitrarily close to $\Gap_f$, the integrality gap of the LP relaxation under function $f(\cdot)$. Examples of such functions are price discounts including percentage discount: $f(x)=ax,a\in[\max_{i,j}\{\frac{p_i}{p_i+q_j},\frac{q_j}{p_i+q_j}\},1)$, and constant discount: $f(x)=x-a,0<a\leq\min\{p_i,q_j\}$. The proof of approximation ratio is similar to that of Theorem \ref{thm:gap-eps}, with the first inequality in \eqref{eq:p+q bound} modified to 
    % \[
    % |f(p_{s,I}+q_{t,J})-R_{IJ}^{st}| \leq 2C\varepsilon r^*,
    % \]
    % and the rest of the proof accordingly. The approximation ratio of Algorithm \ref{alg:gap-eps} is $\Gap_f-4C\varepsilon$.
\end{remark}

\section{Extensions}\label{sec:ext}
In this section, we investigate various extensions of the assortment optimization problem \eqref{prob:IP} by considering assortment capacities, more general bundle revenues, and more categories. Unfortunately, it turns out that all the corresponding assortment optimization problems do not admit approximation algorithms with constant approximation ratio assuming the Exponential Time Hypothesis (ETH). We first introduce the definition of the Bipartite Densest $\kappa$-Subgraph problem (\textsc{BD$\kappa$S}), an analog of the classical Densest $\kappa$-subgraph problem, and the definition of ETH, a widely-used hypothesis in theoretical computer science literature. Assuming ETH, we show that the \textsc{BD$\kappa$S} cannot be approximated with a constant ratio. 
Then, we construct \textit{gap-preserving reductions} from the assortment optimization problem of the two-category MVMNL model with a capacity constraint, and the assortment optimization problem of the two-category MVMNL model with general bundle prices, to \textsc{BD$\kappa$S}. 
Gap-preserving reduction is a commonly used concept when proving hardness of approximation. For a complete definition and more examples of it, one can refer to \cite{williamson2011design}.
Finally, we construct a gap-preserving reduction from the assortment optimization problem of the three-category MVMNL model to the assortment optimization problem of the two-category MVMNL model with general bundle prices. Thus, all three extensions we consider in this section do not have a constant-factor approximation.

Before presenting the hardness results, we formally present the \textsc{BD$\kappa$S} and ETH.

\begin{definition}[Bipartite Densest $\kappa$-Subgraph]
Given an undirected bipartite graph $G=(\textbf{N},\textbf{M},E)$ with a vertex set $\textbf{N}\cup \textbf{M}$ and an edge set $E\subseteq \textbf{N}\times \textbf{M}$, the bipartite densest $\kappa$-subgraph (\textsc{BD$\kappa$S}) problem is defined by
\begin{equation} \label{bdks}
\begin{aligned}
    \max &  \ |E(\textbf{N}_1\times \textbf{M}_1)|  \\
     \text{s.t.} &\  |\textbf{N}_1|=|\textbf{M}_1| = \kappa,\ \textbf{N}_1 \subseteq \textbf{N},\  \textbf{M}_1 \subseteq \textbf{M},
\end{aligned}
\end{equation}
where $E(\textbf{N}_1\times \textbf{M}_1)$ is the set of all edges among the vertices in $\textbf{N}_1\times \textbf{M}_1$.
\end{definition}

\begin{hypothesis}[Exponential Time Hypothesis]
No $2^{o(m)}$-time algorithm can decide whether any 3SAT formula with $m$ clauses is satisfiable.
\end{hypothesis}

ETH, proposed in \cite{impagliazzo2001complexity}, is a commonly used hypothesis in theoretical computer science literature. Although unproven, it is a popular conjecture that is used as the base for analyzing the hardness of various problems. For example, \cite{manurangsi2017almost} analyzed the hardness of the densest-$\kappa$-subgraph problem based on ETH.

The following lemma shows that there is no constant-factor approximation algorithm for \textsc{BD$\kappa$S} under ETH, which will be used to prove inapproximability results of the assortment optimization problems mentioned above.
\begin{lemma}\label{lem:bdks hardness}
There is a constant $c>0$ such that, assuming ETH, there is no $\Omega(g^{-1 /(\log \log g)^{c}})$-approximation algorithm for \textsc{BD$\kappa$S} defined on a bipartite graph with $g$ vertices.
\end{lemma}
\proof{Proof of Lemma \ref{lem:bdks hardness}.}
Given an undirected bipartite graph $G=(\textbf{N},\textbf{M},E)$ with $g=|\textbf{N}\cup \textbf{M}|$, we consider a variant of the \textsc{BD$\kappa$S} as
\begin{equation} \label{dks2}
\begin{aligned}
    \max &\ |E(\textbf{N}_1\times \textbf{M}_1)|  \\
    \text{s.t.} &\  |\textbf{N}_1|+ |\textbf{M}_1| = 2\kappa,\ \textbf{N}_1 \subseteq \textbf{N},\ \textbf{M}_1 \subseteq \textbf{M}.
\end{aligned}
\end{equation}
By Lemma A.3 of \cite{bhaskara2010Detecting} and Theorem 1 of \cite{manurangsi2017almost}, there does not exist a $\Omega(f(g/2))$-approximation algorithm for problem \eqref{dks2}, where  $f(g)=g^{-1 /(\log \log g)^{c}}$ for some constant $c>0$. Suppose there exists a $\Omega(f(g))$-approximation algorithm for \textsc{BD$\kappa$S} defined on $G$. Let $(\textbf{N}_1',\textbf{M}_1')$ be the solution given by this algorithm and $E^*$ be the optimal objective value of \textsc{BD$\kappa$S} on $G$. Clearly, this solution is also feasible in \eqref{dks2}. Let $(\textbf{N}_2,\textbf{M}_2)$ be an optimal solution of \eqref{dks2}. If $|\textbf{N}_2|\neq \kappa$, without loss of generality, assume $|\textbf{N}_2|>\kappa$. Arbitrarily decompose $\textbf{N}_2$  into $\textbf{N}_2',\textbf{N}_2''$ with $|\textbf{N}_2'|=\kappa$. It is clear that $|E(\textbf{N}_2\times \textbf{M}_2)|=|E(\textbf{N}_2'\times \textbf{M}_2)|+|E(\textbf{N}_2''\times \textbf{M}_2)|\leq 2E^*$.
 Since $|E(\textbf{N}_1'\times \textbf{M}_1')|\geq \Omega(f(g))E^*\geq \Omega(f(g/2))E^*$, it follows that $(\textbf{N}_1', \textbf{M}_1')$ is a  $\Omega(f(g/2))$-approximation of problem \eqref{dks2}, which is a contradiction.
\Halmos
\endproof

\subsection{Two-category Capacitated Assortment Optimization} 
The two-category capacitated assortment optimization problem is defined by
\begin{align}\label{prob:cap_two_category}
\begin{aligned}
\max\ & \pi(\textbf{N}_1,\textbf{M}_1)\\
\text{s.t. } & \textbf{N}_1\subseteq \textbf{N},\ \textbf{M}_1\subseteq \textbf{M},\ |\textbf{N}_1|\leq K_1,\ |\textbf{M}_1|\leq K_2,
\end{aligned}
\end{align}
where $\pi(\textbf{N}_1,\textbf{M}_1)$ is the expected revenue of assortment $(\textbf{N}_1,\textbf{M}_1)$ and $K_1,K_2$ are the capacities for category 1 and category 2, respectively. We show that even for equal capacity $K_1=K_2$, there is no constant-factor approximation algorithm for problem \eqref{prob:cap_two_category}.

\begin{theorem}\label{thm:cap_two_category}
Assuming ETH, there does not exist a constant-factor  approximation algorithm for problem \eqref{prob:cap_two_category} with $K_1=K_2=\kappa$.
\end{theorem}
\proof{Proof of Theorem \ref{thm:cap_two_category}.}
We construct an approximation-preserving reduction from \textsc{BD$\kappa$S}. 
Consider an instance of \textsc{BD$\kappa$S} with a bipartite graph $G=(\textbf{N},\textbf{M},E)$ and capacity $\kappa$.
An instance of problem \eqref{prob:cap_two_category} is constructed as follows. Let $\textbf{N}$ and $\textbf{M}$ be the set of products in categories 1 and 2, respectively. Denote $g=|\textbf{N}|+|\textbf{M}|$ the number of vertices in $G$. Let $p_i=q_j=\frac{1}{2}$, $u_{0 j}=u_{i 0}=0$ for $i\in \textbf{N}$, $j\in \textbf{M}$, and
 $u_{i j}=g^{-3}$ for $(i,j)\in E(\textbf{N}\times \textbf{M})$ and $u_{ij}=0$ otherwise. For any $\textbf{N}_1\subseteq \textbf{N},\textbf{M}_1\subseteq \textbf{M}$, it holds that $1 \le 1+  \frac{|E(\textbf{N}_1\times \textbf{M}_1)|}{g^3}\le 1+ \frac{|\textbf{N}_1||\textbf{M}_1|}{g^3} \le 2$.
It follows that the optimal objective of problem \eqref{prob:cap_two_category} satisfies
\begin{equation} \label{eq:cap1}
    \begin{aligned}
        \max_{|\textbf{N}_1|\leq \kappa, |\textbf{M}_1|\leq \kappa}
        \frac{|E(\textbf{N}_1\times \textbf{M}_1)|/g^3}{1+
       |E(\textbf{N}_1\times \textbf{M}_1)|/g^3}\geq \frac{1}{2g^3}\max_{|\textbf{N}_1|= \kappa, |\textbf{M}_1|= \kappa}
        |E(\textbf{N}_1\times \textbf{M}_1)|
    \end{aligned}
\end{equation}
For any feasible solution $(\textbf{N}_1,\textbf{M}_1)$ of problem \eqref{prob:cap_two_category}, we can arbitrarily construct a $(\textbf{N}_2,\textbf{M}_2)$ such that $\textbf{N}_1\subseteq \textbf{N}_2, \textbf{M}_1\subseteq \textbf{M}_2,|\textbf{N}_2|=|\textbf{M}_2|=\kappa$. It holds that
\begin{equation}\label{eq:cap2}
     \pi(\textbf{N}_1,\textbf{M}_1)= \frac{|E(\textbf{N}_1\times \textbf{M}_1)|/g^3}{1+
       |E(\textbf{N}_1\times \textbf{M}_1)|/g^3}\le |E(\textbf{N}_1\times \textbf{M}_1)|/g^3\leq |E(\textbf{N}_2\times \textbf{M}_2)|/g^3.
\end{equation}
By Lemma \ref{lem:bdks hardness} and inequalities \eqref{eq:cap1}\eqref{eq:cap2}, there is no $\Omega(g^{-1 /(\log \log g)^{c}})$-approximation algorithm for problem \eqref{prob:cap_two_category}.
\Halmos
\endproof

\begin{remark}
Theorem \ref{thm:cap_two_category} shows that our problem \eqref{prob:IP} is  harder than the problem studied in \cite{ghuge2022constrained} in some sense. Their cardinality-constrained assortment optimization problem under PCL still has a constant-factor approximation algorithm, while the cardinality-constrained assortment optimization problem under the two-category MVMNL model does not.
\end{remark}

\subsection{Two-category assortment optimization with general bundle prices}\label{subsec:general price}
The model in Section \ref{sec:model} assumes that the price of a bundle is the sum of prices of products in this bundle. In this subsection, we consider a similar two-category choice model but with general bundle prices. Specifically, we assume that the price of bundle $(i,j)$ is $r_{ij}\in \RR_+$ for $i\in \textbf{N}_+,j\in \textbf{M}_+$, where $r_{i0},r_{0j}$ represent the prices of product $i\in \textbf{N},j\in \textbf{M}$, respectively, and $r_{00}=0$. The assortment optimization problem is 
\begin{align}\label{prob:gen_price}
\max_{\textbf{N}_1\subseteq \textbf{N},\textbf{M}_1\subseteq \textbf{M}} \pi(\textbf{N}_1,\textbf{M}_1)=\frac{\sum_{i\in  \textbf{N}_{1+} ,j\in \textbf{M}_{1+} }u_{ij}r_{ij}}{\sum_{i\in \textbf{N}_{1+},j\in \textbf{M}_{1+}}u_{ij}}.
\end{align}

It turns out that there does not exist a constant-factor approximation algorithm for problem \eqref{prob:gen_price}.
\begin{theorem}\label{thm:gen_price}
Assuming ETH, there does not exist a constant-factor  approximation algorithm for problem \eqref{prob:gen_price}.
\end{theorem}

\proof{Proof of Theorem \ref{thm:gen_price}.}

We construct an approximation-preserving reduction from \textsc{BD$\kappa$S}. 
Consider an instance of \textsc{BD$\kappa$S} with a bipartite graph $G=(\textbf{N},\textbf{M},E)$ and capacity $\kappa$.
An instance of problem \eqref{prob:gen_price} is constructed as follows. Let $\textbf{N}$ and $\textbf{M}$ be the set of products in categories 1 and 2, respectively.  Let $r_{i0}=r_{0j}=0$,  $u_{ i 0}= u_{0 j}=\frac{1}{\kappa}$,
and $u_{i j}=\frac{1}{\kappa^2}$  for $i\in \textbf{N}$, $j\in \textbf{M}$. Let 
 $r_{i j}=1$ for $(i,j)\in E(\textbf{N}\times \textbf{M})$ and $r_{ij}=0$ otherwise. Let  $\pi^*$ be the optimal objective value of problem \eqref{prob:gen_price} under this instance.  It is clear that the optimal objective of the \textsc{BD$\kappa$S} is
 \begin{align*}
 	4\max_{|\textbf{N}_1|=|\textbf{M}_1|= \kappa}
        \frac{|E(\textbf{N}_1\times \textbf{M}_1)|}{1+
        \frac{|\textbf{N}_1|+|\textbf{M}_1|}{\kappa}+
        \frac{|\textbf{N}_1||\textbf{M}_1|}{\kappa^2}}\leq \kappa^2\pi^*.
 \end{align*}
 
In the following, for any feasible solution $(\textbf{N}_1,\textbf{M}_1)$ of problem \eqref{prob:gen_price}, we construct a feasible solution $(\textbf{N}_2,\textbf{M}_2)$ of problem \textsc{BD$\kappa$S} in polynomial time such that $\kappa^2\pi(\textbf{N}_1,\textbf{M}_1)\leq c|E(\textbf{N}_2\times \textbf{M}_2)|$ for some constant  $c>0$. Then, Lemma \ref{lem:bdks hardness} and the above inequalities yield the desired result.
We consider three cases below.
\begin{enumerate}[label=(\alph*)]
	\item Suppose $|\textbf{N}_1|\leq \kappa$ and $|\textbf{M}_1|\leq \kappa$. Let $(\textbf{N}_2,\textbf{M}_2)$ be any feasible solution of \textsc{BD$\kappa$S} such that $\textbf{N}_1\subseteq \textbf{N}_2,\textbf{M}_1\subseteq \textbf{M}_2$. We have
	\begin{align}
	\pi(\textbf{N}_1,\textbf{M}_1)=
        \frac{|E(\textbf{N}_1\times \textbf{M}_1)|/\kappa^2}{1+
        \frac{|\textbf{N}_1|+|\textbf{M}_1|}{\kappa}+
        \frac{|\textbf{N}_1||\textbf{M}_1|}{\kappa^2}}\leq 
        |E(\textbf{N}_2\times \textbf{M}_2)|/\kappa^2
	\end{align}
\item Suppose exactly one of $\textbf{N}_1,\textbf{M}_1$ has more than $\kappa$ products. Without loss of generality, assume $|\textbf{N}_1|>\kappa\geq |\textbf{M}_1|$. We randomly choose a subset $\textbf{N}_1'$ of $\textbf{N}_1$ that has exactly $\kappa$ products with equal probabilities. It follows that
\begin{equation}
    \begin{aligned}
        \EE\left(\frac{|E(\textbf{N}_1'\times \textbf{M}_1)|}{\kappa^2}\right)=
        \frac{|E(\textbf{N}_1\times \textbf{M}_1)|}{\kappa^2} \cdot \frac{\kappa}{|\textbf{N}_1|}. 
    \end{aligned}
\end{equation}
Thus, there exists $\textbf{N}_1''\subset \textbf{N}_1$ with $|\textbf{N}_1''|=\kappa$ such that
\begin{align*}
	\pi(\textbf{N}_1'',\textbf{M}_1) = & \frac{|E(\textbf{N}_1''\times \textbf{M}_1)|/\kappa^2}{1+\frac{|\textbf{N}_1''|+|\textbf{M}_1|}{\kappa}+\frac{|\textbf{N}_1''||\textbf{M}_1|}{\kappa^2}}= \frac{1}{2}\frac{|E(\textbf{N}_1''\times \textbf{M}_1)|/\kappa^2}{1+\frac{|\textbf{M}_1|}{\kappa}}\\
	\geq & \frac{1}{2}\frac{|E(\textbf{N}_1\times \textbf{M}_1)|/\kappa^2}{1+\frac{|\textbf{M}_1|}{\kappa}}\frac{\kappa}{|\textbf{N}_1|}\geq \frac{1}{2}\pi(\textbf{N}_1,\textbf{M}_1).
\end{align*}
One such $\textbf{N}_1''$ can be found in polynomial time by greedily pruning vertices with smallest number of neighbors. It then reduces to case (a).
\item Suppose $|\textbf{N}_1|>\kappa$ and $|\textbf{M}_1|> \kappa$. Following a similar approach in (b), we randomly choose a subset $\textbf{N}_1'$ of $\textbf{N}_1$ that has exactly $\kappa$ products and a subset $\textbf{M}_1'$ of $\textbf{M}_1$ that has exactly $\kappa$ products, both with equal probabilities. It follows that
\begin{equation}
    \begin{aligned}
        \EE\left(\frac{|E(\textbf{N}_1'\times \textbf{M}_1')|}{\kappa^2}\right)=
        \frac{|E(\textbf{N}_1\times \textbf{M}_1)|}{\kappa^2} \cdot \frac{\kappa^2}{|\textbf{N}_1||\textbf{M}_1|}. 
    \end{aligned}
\end{equation}
Thus, there exists $\textbf{N}_1''\subset \textbf{N}_1,\textbf{M}_1''\subseteq \textbf{M}_1$ with $|\textbf{N}_1''|=|\textbf{M}_1''|=\kappa$ such that
\begin{align*}
	\pi(\textbf{N}_1'',\textbf{M}_1'') = & \frac{|E(\textbf{N}_1''\times \textbf{M}_1'')|}{4\kappa^2}\geq \frac{1}{4}\frac{|E(\textbf{N}_1\times \textbf{M}_1)|/\kappa^2}{\frac{|\textbf{N}_1||\textbf{M}_1|}{\kappa^2}}\geq \frac{1}{4}\pi(\textbf{N}_1,\textbf{M}_1).
\end{align*}
Again, such $(\textbf{N}_1'',\textbf{M}_1'')$ can be found in polynomial time by greedily pruning vertices with smallest number of neighbors. It then reduces to case (a).
\end{enumerate}

\Halmos
\endproof

\begin{remark}
	From the proof of Theorem \ref{thm:gen_price}, the instance of problem \eqref{prob:gen_price} reduced from \textsc{BD$\kappa$S} satisfies $r_{ij}\geq r_{i0}+r_{0j}$. Some special practical cases of $r_{ij}\leq r_{i0}+r_{0j}$ admit constant-factor approximation algorithms as mentioned in Remark \ref{remark:general price} in Section \ref{subsec:gap-e}. 
\end{remark}

\subsection{Three-category choice model and assortment optimization}
In this subsection, we consider a choice model with three categories and investigate its assortment optimization problem.

Let $\textbf{N},\textbf{M},\textbf{L}$ be the sets of products in categories 1,2,3, respectively. Let $u_{ijl},r_{ijl}\in \RR_+$ be the preference weight and price of bundle $(i,j,l)\in \textbf{N}_+\times \textbf{M}_+\times \textbf{L}_+$, respectively, where $u_{000}=1$ and $r_{ijl}=r_{i00}+r_{0j0}+r_{00l}$ for all $(i,j,l)\in \textbf{N}_+\times \textbf{M}_+\times \textbf{L}_+$. The assortment optimization problem is
\begin{align}\label{prob:3cat}
	\max_{\textbf{N}_1\subseteq \textbf{N},\textbf{M}_1\subseteq \textbf{M}, \textbf{L}_1\subseteq \textbf{L}} &\ \pi(\textbf{N}_1,\textbf{M}_1,\textbf{L}_1)=\frac{\sum_{i\in \textbf{N}_{1+},j\in \textbf{M}_{1+},l\in \textbf{L}_{1+}}r_{ijl}u_{ijl}}{\sum_{i\in \textbf{N}_{1+},j\in \textbf{M}_{1+},l\in \textbf{L}_{1+}}u_{ijl}}.
\end{align}

It turns out that there does not exist a constant-factor approximation algorithm for problem \eqref{prob:3cat}.

\begin{theorem}\label{thm:3cat}
Assuming ETH, there does not exist a constant-factor  approximation algorithm for problem \eqref{prob:3cat}.
\end{theorem}

\proof{Proof of Theorem \ref{thm:3cat}.}
We construct an approximation-preserving reduction from problem  \eqref{prob:gen_price}.

Given an instance of the two-category problem \eqref{prob:gen_price} with price $r_{ij}$ and preference weight $u_{ij}$, we construct a three-category instance as follows. The sets of products of the first two categories remain $\textbf{N},\textbf{M}$, respectively.  In the third category, for each pair $(i,j)\in \textbf{N}_+\times \textbf{M}_+\setminus \{(0,0)\}$, we construct a product $l=(i,j)$ such that $u_{i,j,(i,j)}=\epsilon u_{ij}$ and $u_{s,t,(i,j)}=0$ for $(s,t)\neq (i,j)$, and $r_{0,0,(i,j)}=r_{ij}/\epsilon$, where $\epsilon$ is a positive number. Let $u_{ij0}=u_{ij}$ and $r_{i00}=r_{0j0}=0$ for all $i\in \textbf{N}_+,\ j\in \textbf{M}_+$. All the other parameters are zero. Let $(\textbf{N}_1,\textbf{M}_1)$ be an optimal solution of the two-category problem \eqref{prob:gen_price} and consider a three-category assortment $(\textbf{N}_1,\textbf{M}_1,\textbf{L}_1)$ with $\textbf{L}_1=\textbf{N}_{1+}\times \textbf{M}_{1+}\setminus \{(0,0)\}$. It is clear that
\begin{align*}
\begin{aligned}
	\pi(\textbf{N}_1,\textbf{M}_1,\textbf{L}_1)=& \frac{\sum_{i\in \textbf{N}_{1+},j\in \textbf{M}_{1+}}r_{ij}u_{ij}}{\sum_{i\in \textbf{N}_{1+},j\in \textbf{M}_{1+}}u_{ij}+\epsilon\sum_{(i,j)\in \textbf{N}_{1+}\times \textbf{M}_{1+}\setminus \{(0,0)\}}u_{ij}}\\
\geq & \frac{1}{1+\epsilon}\frac{\sum_{i\in \textbf{N}_{1+},j\in \textbf{M}_{1+}}r_{ij}u_{ij}}{\sum_{i\in \textbf{N}_{1+},j\in \textbf{M}_{1+}}u_{ij}}=\frac{1}{1+\epsilon}\pi(\textbf{N}_1,\textbf{M}_1).
\end{aligned}
\end{align*}
Next, for any three-category assortment $(\textbf{N}_2,\textbf{M}_2,\textbf{L}_2)$, we show that $\pi(\textbf{N}_2,\textbf{M}_2,\textbf{L}_2)\leq \pi(\textbf{N}_2,\textbf{M}_2)$. Theorem \ref{thm:gen_price} and the above inequalities then yield the desired result.

Note that the expected revenue does not change if we remove all product $(s,t)\in \textbf{L}_2$ such that $(s,t)\notin \textbf{N}_{2+}\times \textbf{M}_{2+}$ by the definition of bundle prices. If there exists $(i,j)\in \textbf{N}_{2+}\times \textbf{M}_{2+}\setminus \{(0,0)\}$ such that $r_{ij}>0$, then we add product $(i,j)$ into $\textbf{L}_2$ if it is not included. Denote the new assortment of the third category by $\textbf{L}_2'$. For sufficiently small $\epsilon$ (e.g., $\epsilon=\min\{r_{ij}\mid r_{ij}> 0, (i,j)\in \textbf{N}_+\times \textbf{M}_+\}/\max\{r_{ij}+1\mid (i,j)\in \textbf{N}_+\times \textbf{M}_+\}$), we have $\pi(\textbf{N}_2,\textbf{M}_2,\textbf{L}_2)\leq \pi(\textbf{N}_2,\textbf{M}_2,\textbf{L}_2')$. Moreover, it holds that
\begin{align*}
	\pi(\textbf{N}_2,\textbf{M}_2,\textbf{L}_2')=& \frac{\sum_{i\in \textbf{N}_{2+},j\in \textbf{M}_{2+}}r_{ij}u_{ij}}{\sum_{i\in \textbf{N}_{2+},j\in \textbf{M}_{2+}}u_{ij}+\sum_{(i,j,l)\in \textbf{N}_{2+}\times \textbf{M}_{2+}\times \textbf{L}_2'}u_{ijl}}\\
\leq & \frac{\sum_{i\in \textbf{N}_{2+},j\in \textbf{M}_{2+}}r_{ij}u_{ij}}{\sum_{i\in \textbf{N}_{2+},j\in \textbf{M}_{2+}}u_{ij}}=\pi(\textbf{N}_2,\textbf{M}_2).
\end{align*}
Thus, we prove that $\pi(\textbf{N}_2,\textbf{M}_2,\textbf{L}_2)\leq \pi(\textbf{N}_2,\textbf{M}_2)$.
\Halmos
\endproof
\begin{remark}
    Theorem \ref{thm:3cat} implies that there is no constant-factor approximation algorithm for MVMNL models with more than two categories. This is in contrast with the result in \cite{lyu2021assortment} where an FPTAS exists for any fixed number of categories under the MVMNL model with group-dependent interactions. Thus, by considering arbitrary product interactions in a bundle, the problem becomes much harder.
\end{remark}

\section{Numerical Experiment}\label{sec:numerical}

In this section, we conduct a numerical experiment to demonstrate the performance of our proposed approximation framework for the assortment optimization problem \eqref{prob:IP}. As we will see from the numerical results, the actual performance of the approximation algorithms is much better than the theoretical approximation guarantee derived in previous sections.

In our numerical experiments, we generate \emph{i.i.d.}~random prices for all individual products from a uniform distribution, and the bundle price is just the sum of individual prices. The preference weights of bundles and products $u_{ij}$ are generated from \emph{i.i.d.}~uniform distributions (we also tried different distribution such as exponential distribution and log-uniform distribution and obtained similar results).
% Different scales of instances are also tested. 
We conduct test on different scales of $n=m\in\{25,50,75,100,150\}$. For each scale, we run $10,000$ replicates with different randomly generated instances. All the numerical experiments are run on the same computer with a Intel Core-i7 cpu and 32 GB of memory, the linear programs are solved by Gurobi version 9.1.1 with Python version 3.7.6.

We compare the performance of the assortments given by three different approximation algorithms proposed in this paper: the assortment $(x^{\ARO},y^{\ARO})$ from Algorithm \ref{alg:half}, the assortment $(x^{K=4},y^{K=4})$ generated by our partition-and-optimize approximation framework with $K=4$, where our theoretical approximation guarantee is $\frac{5+\sqrt{5}}{10}\approx 0.7236$, and the assortment $(x^\varepsilon,y^\varepsilon)$ from Algorithm \ref{alg:gap-eps} with $\varepsilon=0.1$. As a benchmark, we consider the independent random rounding of the LP optimal solution, namely, round $\Tilde{x}^{\mathrm{LP}}=0.5$ to 0 or 1 with equal probability (similar for $\Tilde{y}^{\mathrm{LP}}=0.5$), and denote the rounded solution as $(x^{rr},y^{rr})$.
%As a benchmark, we utilize a common heuristic of revenue-ordered assortments and find the best assortment that is revenue-ordered in both product categories by enumeration, denoted as $(x_{\mathrm{RO}},y_{\mathrm{RO}})$.  

Due to limited computational power, we are not able to obtain the true optimal revenue $\pi^*$ for the original problem \eqref{prob:IP} for the scale of problems in the numerical experiments. Thus, we use the optimal objective value $r^*$ of the LP relaxation \eqref{prob:LP} as an upper bound of $\pi^*$. As we shall observe in the numerical study, the probability that $r^*=\pi^*$ is high in the randomly-generated instances. Note that when $r^*=\pi^*$, there exists an optimal solution of the LP relaxation \eqref{prob:LP} where all $(\Tilde{x}^{\mathrm{LP}},\Tilde{y}^{\mathrm{LP}},\Tilde{z}^{\mathrm{LP}})$ are integral and is a feasible solution to the assortment optimization problem \eqref{prob:IP}.
%To obtain the true optimal solution for the original problem \eqref{prob:IP}, we enumerate for $n=m=10$, for other problem scales, we use the optimal objective value of the LP relaxation \eqref{prob:LP} as an upper bound of the optimal objective value, due to limited computation power. To illustrate the performance of our proposed approximation framework, we implemented the approximation framework under $K=4$, where our theoretical approximation guarantee is $\frac{5+\sqrt{5}}{10}\approx 0.7236$. 

% Since the original problem is a maximization problem,
We measure the approximation ratio of an assortment $(x,y)$ by  the following formula:
\begin{equation*}
    \alpha(x,y) = \frac{\pi(x,y)}{r^*}\in [0,1].
\end{equation*}
% For each algorithm and problem scale, we compute $\Bar{\alpha}(x,y)$, the average value of $\alpha(x,y)$. 
To measure the average approximation performance of the LP-based algorithms, we compute the average of $\alpha(x^{K=4},y^{K=4})$,  $\alpha(x^\varepsilon,y^\varepsilon)$ and $\alpha(x^{rr},y^{rr})$ over all instances with $r^*\neq\pi^*$, i.e. the LP relaxation's optimal solution is non-integral.
% For $\Bar{\alpha}(x^{K=4},y^{K=4})$,  $\Bar{\alpha}(x^\varepsilon,y^\varepsilon)$ and $\Bar{\alpha}(x^{rr},y^{rr})$, we calculate the average over all instances where $r^*\neq\pi^*$, i.e. the LP relaxation's optimal solution is non-integral. 
% If we compute $\Bar{\alpha}(x^{K=4},y^{K=4})$,  $\Bar{\alpha}(x^\varepsilon,y^\varepsilon)$ on all random instances, the values of $\Bar{\alpha}(x^{K=4},y^{K=4})$ and $\Bar{\alpha}(x^\varepsilon,y^\varepsilon)$ will be larger than $0.9999$ in all entries. 
The average of $\alpha(x^{\ARO},y^{\ARO})$ is computed over all $10,000$ instances. For each algorithm, denote $\Bar{\alpha}(x,y)$ the average approximation performance.  We also report the worst-case performance of the algorithms found in all $10,000$ random instances, denoted by $\min \{\alpha(x,y)\}$, as well as the average run-time per instance in seconds. We report the results of the numerical experiment in Table \ref{table:numerical}.

\begin{table}[htbp]
\centering
\begin{tabular}{l|rrrrr}
\hline
                  $m\times n$                                                   & $25\times 25$ & $50\times 50$ & $75\times 75$ & $100\times 100$ & $150\times 150$ \\ \hline
$\mathbb{P}[r^*\neq \pi^*]$ & 1.19\%        & 2.78\%         & 5.11\%        & 6.82\%        & 9.97\%        \\ \hline
$\Bar{\alpha}(x^{K=4},y^{K=4})$                                                     & 0.9960        & 0.9974        & 0.9981        & 0.9984        & 0.9988        \\
$\min\{\alpha(x^{K=4},y^{K=4})\}$                                        & 0.9729        & 0.9885        & 0.9890        & 0.9907    & 0.9939    \\
average run-time (seconds)                                       & 0.0325        & 0.1440        & 0.3812        & 0.8017       & 2.2431      \\ \hline
$\Bar{\alpha}(x^\varepsilon,y^\varepsilon)$                                        & 0.9976        & 0.9979        & 0.9990        & 0.9991        & 0.9993        \\
$\min\{\alpha(x^\varepsilon,y^\varepsilon)\}$                                          & 0.9923        & 0.9975        & 0.9930        & 0.9823        & 0.9956        \\
average run time (seconds)                                        & 0.0332        & 0.1489        & 0.3845        & 0.8156       & 2.2875      \\ \hline
$\Bar{\alpha}(x^{\ARO},y^{\ARO})$                                                   & 0.8983        & 0.8832        & 0.8716        & 0.8729        & 0.8675        \\
$\min\{\alpha(x^{\ARO},y^{\ARO})\}$                                                  & 0.5162        & 0.5392        & 0.5295        & 0.5310         & 0.5354        \\
average run time (seconds)                            & 0.0226        & 0.0892      & 0.2036      & 0.3562       & 0.5587       \\ \hline
$\Bar{\alpha}(x^{rr},y^{rr})$                                                   & 0.9334        & 0.9508        & 0.9579        & 0.9598        & 0.9566        \\
$\min\{\alpha(x^{rr},y^{rr})\}$                                                  & 0.5785        & 0.5344        & 0.7983        & 0.7358         & 0.8627        \\
average run time (seconds)                            & 0.0325        & 0.1439      & 0.3810      & 0.7998       & 2.2425       \\

\hline
\end{tabular}
\caption{Results of numerical experiment}\label{table:numerical}
\end{table}

From Table \ref{table:numerical}, we have the following observations. 
First, the performance of our partition-and-optimize approximation framework with $K=4$ on the randomly-generated problem instances is much better than the theoretical approximation guarantee. The performances of $(x^{K=4},y^{K=4})$ and $(x^\varepsilon,y^\varepsilon)$ are similar in terms of the approximation ratio and run time. Both are significantly better compared to the benchmark random-rounding method in terms of average and worst-case performance. We can also observe from the experiment, as the scale of the problem instance increases, the average performance and the worst performance of our proposed algorithms in all random instances both increase. This indicates that our approximation framework scales well into large problem sizes.
Second, the proportion of random instances where the LP relaxation does not give an integral $(\Tilde{x}^{\mathrm{LP}},\Tilde{y}^{\mathrm{LP}},\Tilde{z}^{\mathrm{LP}})$ is small, indicating that in most of the random instances, the LP relaxation \eqref{prob:LP} is effective in solving the assortment optimization problem. Third, Algorithm \ref{alg:half} with a $0.5$-approximation ratio also performs far better than the theoretical approximation guarantee in terms of average performance, though it performs significantly worse compared to the other two methods, in terms of both average and worst-case performance.
%However, we observe in our numerical experiments, that as the scale of the problem increases, the probability that the LP relaxation does not give an integral $(\Tilde{x}^{\mathrm{LP}},\Tilde{y}^{\mathrm{LP}},\Tilde{z}^{\mathrm{LP}})$ increases.
\section{Concluding remarks}

In this paper, we studied the assortment optimization problem considering customers' multi-purchase behavior based on the two-category MVMNL model. Based on an LP relaxation, we develop an approximation framework tailored to this problem that achieves a 0.74 approximation guarantee and show how to close the integrality gap of the LP relaxation. We prove the hardness of extending our framework into more general settings. Our work is among the first few papers that consider the assortment optimization problem under the \textit{multi-choice} models, and the first paper to derive approximation algorithms with constant approximation ratios, while allowing arbitrary interaction structure between products in a bundle. 

Customers' \textit{multi-choice} behavior is ubiquitous in the business world, where the purchase decision of multiple products are made jointly. Thus, understanding the assortment optimization problem related to \textit{multi-choice} models are practically important. However, as we show in the paper, the assortment optimization problem for the two-category MVMNL model allowing arbitrary interactions between products is strongly NP-hard. Its extensions into more general settings are even harder. So, an interesting future direction would be identifying and exploiting the structure of the product interactions to tackle the related assortment optimization problem, or analyze the assortment optimization problem with bundle utilities drawn from a random distribution. 

%We note that although the problem of solving the assortment optimization for the multivariate MNL model with more categories is theoretically hard, it is of practical interest to develop efficient algorithms to achieve an acceptable performance. Thus, analyzing the performance of the algorithm under certain interactions or random utility distributions might be an important future direction.

% Appendix here
% Options are (1) APPENDIX (with or without general title) or
%             (2) APPENDICES (if it has more than one unrelated sections)
% Outcomment the appropriate case if necessary
%
% \begin{APPENDIX}{<Title of the Appendix>}
% \end{APPENDIX}
%
%   or
%
\begin{APPENDICES}
\section{Proof of Theorem \ref{thm:NPhard}.}\label{Appendix:proof}
We construct a polynomial-time reduction from the \textsc{Max-DiCut} problem on a directed acyclic graph (DAG). The decision problem of \textsc{Max-DiCut} on DAG consists of a DAG  $G=(V,E)$ with a vertice set $V=\{v_1,...,v_n\}$ and an edge set $E$, a weight $w_{ij}\in \ZZ_{++}$ of each edge $(i,j)\in E$, and a positive number $t$. The problem asks whether there exists a vertex subset with indicator variable $x\in \{0,1\}^n$ such that the total weight of outgoing edges is no less than $t$, i.e.,
\begin{align*}
     \text{whether } c^*\overset{\underset{\mathrm{def}}{}}{=}  \max_{x\in\{0,1\}^n}\sum_{(i,j)\in E}w_{ij}x_i(1-x_j)\geq t?
\end{align*}

We construct an instance of problem \eqref{prob:IP} as follows.
Without loss of generality, assume $v_1,...,v_n$ is the topological ordering of $G$, i.e., any directed edge $(i,j)$ of $E$ satisfies $i<j$.
Scale all $w_{ij}$ and $t$ simultaneously such that $t\geq 2n$. Let the set of products in categories 1 and 2 be $\textbf{N}$ and $\textbf{M}$ respectively, and denote the products in $\textbf{N}$ (as well as the ones in $\textbf{M}$) by the numbers  $\{1, 2, \dots, n\}$. Let $p_i=t-0.5-i$ and $q_j=j$ for $i\in \textbf{N},j\in \textbf{M}$. Let $u_{ii}=2(s+1)$ for each $i\in \textbf{N}$, where $s=\sum_{(i,j)\in E}w_{ij}$. Let $u_{ij}=\frac{w_{ij}}{j-i-0.5}$, if $1\leq i<j\leq n$ and $(i,j)\in E$, and $u_{ij}=0$ otherwise. Denote the optimal objective value under this instance by $\pi^*$.

We prove that
\begin{align}
   \pi^*\geq t \text{ if and only if } c^*\geq t .
\end{align}
It is easy to see that $\pi^*\geq  t$ is equivalent to $g(t)\geq t$, where $g(t)$ is defined by
\begin{align}
g(t)=&\max\left\{g(x,y,t)\mid x,y\in\{0,1\}^n\right\}, \label{eq:gap_reduction_3}\\
g(x,y,t)=&\sum_{i\in \mathbf{N},j\in \mathbf{M}} u_{ij}(p_i+q_j-t)x_iy_j+\sum_{i\in \mathbf{N}}u_{i0}(p_i-t)x_i+\sum_{j\in \mathbf{M}}u_{0j}(q_j-t)y_j.\notag
\end{align}
It is cleat that $g(t)\geq 0$ as $g(\mathbf{0,0},t)=0$ for any $t$.
Plug all the parameter values into $g(x,y,t)$, we have
\begin{align*}
    g(x,y,t)&=  \sum_{(i,j)\in E}\frac{w_{ij}}{j-i-0.5}(t-0.5-i+j-t)x_iy_j-\sum_{i=1}^n(s+1)x_iy_i\\
    & = \sum_{(i,j)\in E}w_{ij}x_iy_j-(s+1)\sum_{i=1}^nx_iy_i
\end{align*}

Let $(x^*,y^*)$ be an optimal solution of problem \eqref{eq:gap_reduction_3}. If $x^*_i=y^*_i=1$ for some $i\in \textbf{N}$, then $f(x^*,y^*)\leq \sum_{(i,j)\in E}w_{ij}-(s+1)=-1<0$. Thus, $x^*_i+y^*_i\leq 1$ for all $i\in \textbf{N}$. If $x^*_k=y^*_k=0$ for some $k\in \textbf{N}$, then
\begin{align*}
    & f(x^*+\mathbf{e}_k,y^*)\\&
    =\sum_{(i,j)\in E}w_{ij}x^*_iy^*_j+\sum_{l:(k,l)\in E}w_{kl}y^*_l-(s+1)\sum_{i\neq k}x^*_iy^*_i\\
    & \geq \sum_{(i,j)\in E}w_{ij}x^*_iy^*_j-(s+1)\sum_{i=1}^nx^*_iy^*_i\\
    & =f(x^*,y^*).
\end{align*}
It implies that there exists an optimal solution such that $x_k^*+y^*_k=1$ for all $k\in \textbf{N}$. Therefore,
$$g(t)=\max_{x\in \{0,1\}^n} \sum_{(i,j)\in E}w_{ij}x_i(1-x_j)=c^*,$$
and thus $c^*\geq t$ if and only if $\pi^*\geq t$.
\Halmos

\end{APPENDICES}

%\theendnotes

% Acknowledgments here
%\ACKNOWLEDGMENT{}

% References here (outcomment the appropriate case)

% CASE 1: BiBTeX used to constantly update the references
%   (while the paper is being written).
\bibliographystyle{informs2014} % outcomment this and next line in Case 1
\bibliography{reference.bib} % if more than one, comma separated

% CASE 2: BiBTeX used to generate mypaper.bbl (to be further fine tuned)
%\input{mypaper.bbl} % outcomment this line in Case 2

%If you don't use BiBTex, you can manually itemize references as shown below.

%% Here starts the e-companion (EC)
%%%%%%%%%%%%%%%%%%%%%%%%%%%%%%%%%%%%%%%%%%%%%%%%%%%%%%%%%%
%\ECSwitch

%\ECDisclaimer
%%%%%%%%%%%%%%%%%%%%%%%%%%%%%%%%%%%%%%%%%%%%%%%%%%%%%%%%%%

%%% Main head for the e-companion
% \ECHead{Electronic companion}

%%%%%%%%%%%%%%%%%
\end{document}